\documentclass[1p,sort&compress]{elsarticle}

\usepackage{times}
\usepackage{amsmath}
\usepackage{amssymb,algorithmic}
\usepackage{color}

\makeatletter
\def\ps@pprintTitle{%
  \let\@oddhead\@empty
  \let\@evenhead\@empty
  \def\@oddfoot{\reset@font\hfil\thepage\hfil}
  \let\@evenfoot\@oddfoot
}
\makeatother

\newcommand{\sk}[1]{#1}

\newcommand{\ignore}[1]{}

\newcommand{\BB}{\vspace*{-\medskipamount}}
\newcommand{\BBB}{\vspace*{-\bigskipamount}}

\newcommand{\FF}{\vspace*{\medskipamount}}

\newcommand{\T}{\hspace*{1em}}

\newtheorem{theorem}{Theorem}[section]
\newtheorem{lemma}[theorem]{Lemma}

\newtheorem{definition}{Definition}[section]
\newenvironment{proof}{{\bf Proof.}}{\hfill$\Box$}

\newcommand{\act}[1]{%
    \relax\ifmmode
        \mathord{\mathcode`\-="702D\sf #1\mathcode`\-="2200}%
    \else
        {\small {\sf #1}}%
    \fi
}

\newcommand{\st}{s}
\newcommand{\stt}{t}
\newcommand{\acts}{\pi}
\newcommand{\EXF}{\alpha}
\newcommand{\EX}{\alpha}

\newcommand{\pset}{\mathcal{P}}
\newcommand{\mset}{\mathcal{X}}
\newcommand{\tset}{\mathcal{J}}

\newcommand{\sv}[1]{x_{#1}}

\newcommand{\ex}[1]{execs\left(#1\right)}

\newcommand{\status}[1]{\text{\sc{status}}_{#1}}
\newcommand{\posA}[1]{\text{\sc{pos}}_{#1}}
\newcommand{\pos}[2]{\text{\sc{pos}}_{#1}\left(#2\right)}
\newcommand{\nxt}[1]{\text{\sc{next}}_{#1}}
\newcommand{\next}[1]{\text{\sc{next}}_{#1}}
\newcommand{\tmp}[1]{\text{\sc{tmp}}_{#1}}
\newcommand{\q}[1]{\text{\sc{q}}_{#1}}

\newcommand{\algName}{\mathrm{KK}_\beta}

\newcommand{\DONE}{\mathrm{DONE}}
\newcommand{\TRY}{\mathrm{TRY}}
\newcommand{\FREE}{\mathrm{FREE}}
\newcommand{\SET}{\mathrm{SET}}
\newcommand{\SuperSET}{\mathrm{SuperSet}}

\newcommand{\Pool}{\mathrm{POOL}}
\newcommand{\Stuck}{\mathrm{STUCK}}

\newcommand{\DONEinS}[2]{#1.\DONE_{#2}}
\newcommand{\TRYinS}[2]{#1.\TRY_{#2}}
\newcommand{\FREEinS}[2]{#1.\FREE_{#2}}

\newcommand{\setA}{\mathrm{A}}
\newcommand{\setB}{\mathrm{B}}

\newcommand{\BigO}[1]{\mathrm{O}(#1)}

\newcommand{\size}[1]{\mathrm{size}_{p,#1}}
\newcommand{\mapf}[1]{\mathrm{map}\left(#1\right)}
\newcommand{\ItAlg}[1]{\mathrm{IterativeKK}\left(#1\right)}
\newcommand{\WItAlg}[1]{\mathrm{WA\_IterativeKK}\left(#1\right)}
\newcommand{\ItStep}[1]{\mathrm{IterStepKK\left(#1\right)}}
\newcommand{\WItStep}[1]{\mathrm{WA\_IterStepKK\left(#1\right)}}
\newcommand{\ItStepName}{\mathrm{IterStepKK}}
\newcommand{\WItStepName}{\mathrm{WA\_IterStepKK}}

\newcommand{\doset}[3]{len\left(#1|_{\act{do}_{#2,#3}}\right)}

\newcommand{\donum}[1]{Do(#1)}

\newcommand{\fe}{{\it fairexecs}}

\newcommand{\Par}[1]{\medskip\noindent{\bf #1}~}

\newcommand{\nn}[1]{#1}

\newcommand{\ip}{immediate predecessor}
\newcommand{\ipSym}{\mapsto}

\newcommand{\rank}[2]{\left[ #1 \right]_{#2}}

\newcommand{\twocolcode}[4]{\BB\noindent
 \begin{minipage}[t]{#1\textwidth}{\footnotesize ~ #3}\end{minipage}
 \begin{minipage}[t]{#2\textwidth}{\footnotesize ~ #4}\end{minipage}
}

\newcommand{\fourcolcode}[8]{\BB
 \begin{minipage}[t]{#1\textwidth}{\footnotesize ~ #5}\end{minipage}
 \begin{minipage}[t]{#2\textwidth}{\footnotesize ~ #6}\end{minipage}
 \begin{minipage}[t]{#3\textwidth}{\footnotesize ~ #7}\end{minipage}
 \begin{minipage}[t]{#4\textwidth}{\footnotesize ~ #8}\end{minipage}
}

\newcommand{\ef}[2]{
\begin{tabbing}
XX\= XX\=XX\=XX\=XX\=XX\=  \kill
\protect #1\\
Effect: \+ \\
  #2 \-
\end{tabbing}}

\newcommand{\prcef}[3]{\begin{tabbing}
X\=XX\=XX\=XX\=XX\= \kill
\protect #1\\
Precondition: \+ \\
#2 \- \\
Effect: \+ \\
#3 \-
\end{tabbing}}

\begin{document}

\begin{frontmatter}

\title{Solving the At-Most-Once Problem  with Nearly Optimal Effectiveness}

\author[UCONN]{Sotirios Kentros\fnref{fn1}}
\ead{skentros@engr.uconn.edu}

\author[UCONN]{Aggelos Kiayias\fnref{fn2}}
\ead{aggelos@kiayias.com}

\address[UCONN]{Computer Science and Engineering, University of Connecticut, 
          			Storrs, USA}

\fntext[fn1]{Research supported in part by the State Scholarships Foundation of
Greece.}
\fntext[fn2]{Research supported in part by NSF awards 0831304, 0831306 
and EU projects RECUP and CODAMODA}
\fntext[fn3]{NOTICE: This is the author's version of a work that was accepted for publication in Theoretical Computer Science. Changes resulting from the publishing process, such as peer review, editing, corrections, structural formatting, and other quality control mechanisms may not be reflected in this document. Changes may have been made to this work since it was submitted for publication. A definitive version was subsequently published in Theoretical Computer Science, [Volume 496, 22 July 2013] DOI:10.1016/j.tcs.2013.04.017}

\begin{abstract}
We present and analyze a wait-free deterministic algorithm for solving the at-most-once problem: how $m$ shared-memory fail-prone
processes perform asynchronously $n$ jobs {\em at most once}.
Our algorithmic strategy provides for the first time 
nearly  optimal  effectiveness, which is a measure that expresses the
total number of jobs completed in the worst case. 
The effectiveness of our algorithm equals $n-2m+2$. This is
 up to an additive  factor of $m$  close to the known effectiveness upper
bound $n-m+1$ over all possible algorithms and improves
on the previously best known  deterministic solutions that have
effectiveness only $n-\log m \cdot \mathrm{o}(n)$.
We also present an iterative version of our algorithm 
that for any  $m = \BigO{\sqrt[3+\epsilon]{n/\log n}}$ 
is both  effectiveness-optimal and
work-optimal, for any constant $\epsilon > 0$.
We then  employ this algorithm  to provide a new 
algorithmic solution for the Write-All problem 
which is work optimal for any $m=\BigO{\sqrt[3+\epsilon]{n/\log n}}$.
\end{abstract}


\end{frontmatter}

\section{Introduction}
\label{sec:intro}
The \emph{at-most-once problem} for asynchronous shared memory systems 
was introduced by Kentros \emph{et al.}\,\cite{KKNS09} as the problem of  performing a set of $n$ jobs by $m$ fail-prone processes 
while maintaining at-most-once semantics. 

The \emph{at-most-once} semantic for object invocation ensures that an operation accessing and altering the state of an object is performed no more than once.
This semantic is among the standard semantics for remote procedure calls (RPC) and method invocations and it provides important
means for reasoning about the safety of critical applications.
Uniprocessor systems may trivially provide solutions for at-most-once
semantics by implementing a central schedule for operations. 
The problem becomes very challenging for autonomous processes
in a  system
with concurrent
invocations on multiple objects.
At-most-once semantics have been thoroughly studied in the context
of at-most-once message delivery\,\cite{CCW95,LLS93,LLW91} 
and
at-most-once process invocation for 
RPC~\cite{BirNel84,LinGan85,Spector82}. 
However, finding effective solutions
for asynchronous shared-memory multiprocessors, in terms of how many
at-most-once invocations can be performed by the cooperating processes,
 is largely an open problem. Solutions for the at-most-once problem, using 
only  atomic read/write memory,
and without specialized hardware support such as conditional writing,
provide a useful tool in reasoning about the safety
properties of applications developed for a variety of 
multiprocessor systems, including those not supporting
bus-interlocking instructions and multi-core systems. 
Specifically, in recent years,  
attention has shifted  from increasing clock speed 
towards \emph{chip multiprocessing}, in order 
to increase the performance of systems. Because
of the differences in each multi-core system, asynchronous shared memory
is becoming an important abstraction for arguing about the safety properties of 
parallel applications in such systems. In the next years, one can expect chip
multiprocessing to appear in a wide range of applications, many of which 
will have components that need to satisfy at-most-once semantics in order to
guarantee safety. Such applications may include autonomous robotic
devices, robotic devices for assisted living, automation in production lines
or medical facilities. In such applications performing specific jobs 
at-most-once may be of paramount importance for safety of patients, the workers in a facility, or the devices themselves. Such jobs could be the triggering of a
motor in a robotic arm, the activation of the X-ray gun in an X-ray machine,
or supplying a dosage of medicine to a patient. 

Perhaps the most important  question in this area is 
devising algorithms for the at-most-once problem with good \emph{effectiveness}.  The complexity measure of  effectiveness\,\cite{KKNS09} describes	the number of jobs completed (at-most-once) by an implementation, as a function of the overall number of jobs $n$, 
the number of processes $m$,  and the number of  crashes $f$. 
The only deterministic solutions known, 
exhibit very low effectiveness
$( n^\frac{1}{\log m} -1)^{\log m}$\,(see\,\cite{KKNS09}) 
which for most choices of the parameters is very far from optimal (unless $m=\BigO{1}$).
Contrary to this, the present work presents the first wait-free deterministic
algorithm for the at-most-once problem 
which is  optimal up to additive factors of $m$.
Specifically our effectiveness is $n-(2m-2)$ which
comes close to an additive factor of $m$ to the known   upper
bound over all possible algorithms for effectiveness  $n-m+1$\,(from~\cite{KKNS09}).
We also demonstrate how to construct an algorithm which has effectiveness  $n-\BigO{m^2 \log n \log m}$ and work complexity 
$\BigO{n +  m^{3+\epsilon} \log n}$, and is both effectiveness and work optimal when $m = \BigO{\sqrt[3+\epsilon]{n / \log n}}$,
for any constant $\epsilon > 0$ (work complexity counts the total
number of basic operations performed by the processes). 
Finally we show how to use this algorithm in order to solve the \emph{Write-All} problem
\cite{Alex97} with work complexity  $\mathrm{O}(n +  m^{3+\epsilon} \log n)$.

\Par{Related Work:}
A wide range of works study 
at-most-once semantics in a variety of  settings.
At-most-once message delivery \cite{CCW95,LLS93,LLW91,Watson89} 
and at-most-once semantics for 
RPC \cite{BirNel84,LinGan85,Argus88,LLW91,Spector82}, 
are two areas that
have attracted a lot of attention. 
Both in at-most-once message delivery and RPCs, we
have two entities (sender/client and receiver/server) that 
communicate by message passing. Any entity
may fail and recover and messages may be delayed or lost.
In the first case one wants to guarantee that duplicate messages
will not be accepted by the receiver, while in the case of RPCs, 
one wants to guarantee that the procedure called in the remote server
will be invoked at-most-once\,\cite{Spector82}. 

In Kentros \emph{et al.}\,\cite{KKNS09}, the at-most-once problem for asynchronous shared memory systems and the correctness 
properties to be satisfied by any solution were defined. The first algorithms that solve the at-most-once
problem were provided and analyzed. Specifically  they presented 
two algorithms that solve the	at-most-once problem for two processes with optimal effectiveness 
and a multi-process algorithm, that employs a two-process algorithm as a building block,
and solves the 	at-most-once problem with effectiveness $n-\log m \cdot \mathrm{o}(n)$ and work complexity
$\mathrm{O}(n + m\log m)$. Subsequently Censor-Hillel\,\cite{Hillel2010} provided a probabilistic algorithm
 in the same setting with optimal effectiveness and expected work complexity $\mathrm{O}(nm^2\log m)$ by 
 employing a probabilistic multi-valued consensus  protocol as a building block. 

Following the conference version of this paper\,\cite{KK12} and motivated by the difficulty of implementing wait-free deterministic
solutions for the at-most-once problem that are effectiveness optimal, 
Kentros\,\emph{et al.}\,\cite{SCA12} introduced the \emph{strong at-most-once problem} and studied its feasibility. The strong at-most-once problem refers to the 
setting where effectiveness is measured only in terms of the jobs that need to be executed and the processes that took part in the computation and crashed. The strong
at-most-once problem demands solutions that are adaptive, in the sense that the 
effectiveness depends only on the behavior of processes that participate in the execution. In this manner trivial solutions are excluded and, as demonstrated in\,\cite{SCA12}, processes have to
solve an agreement primitive in order to make progress and provide a solution for the problem. Kentros \emph{et al.}\,\cite{SCA12} prove that the strong at-most-once
problem has consensus number $2$ as defined by Herlihy~\cite{MH91} and
 observe that it belongs in the \emph{Common$2$} class as defined by Afek~\emph{et al.}\,\cite{Afek93}.
As a result, there
exists no wait-free deterministic solution for the strong at-most-once problem
in the asynchronous shared memory model, using atomic read/write registers. 
Kentros \emph{et al.}\,\cite{SCA12} present a randomized $k$-adaptive effectiveness optimal solution 
for the strong at-most-once problem, with expected work complexity of $\BigO{n + k^{2+\epsilon}\log n}$ for any small constant $\epsilon$, where $k$ the number of processes that participate in the execution.

Di Crescenzo and Kiayias in\,\cite{GK05} (and later Fitzi \emph{et al.}\,\cite{Fitzi07}) demonstrate the use 
of the at-most-once semantic in message passing systems for the purpose of secure communication.
Driven by the
fundamental security requirements of {\em one-time pad} encryption,
the authors partition a common random pad among multiple communicating 
parties. Perfect security can be achieved only if 
every piece of the pad is used at most once. 
The authors show how the parties maintain  security while maximizing efficiency 
by applying  at-most-once semantics on pad expenditure. 

Ducker~\emph{et al.}\,\cite{Rotem12} consider a distributed task allocation problem,
where players that communicate using a shared blackboard or an arbitrary directed
communication graph, want to assign the tasks so that each task is performed exactly
once. They consider synchronous execution without failures and examine the communication
and round complexity required to solve the problem, providing relevant lower 
and upper bounds. If crashes are introduced in their model,  
the impossibility results from Kentros ~\emph{et al.}\,\cite{KKNS09} will apply
to the at-most-once version of their problem.

Another related problem is the semi-matching problem\,\cite{Harvey06, Bokal12, Czygrinow12}. 
The semi-matching problem known also as the load balancing problem has been
extensively studied under various names in the network scheduling literature. 
Recently it has received renewed attention after a paper by Harvey~\emph{et al.}\,\cite{Harvey06}, where the name semi-matching was introduced. Semi-matching can be seen as an abstraction of 
the problem of matching clients with servers, each of which can process a subset of clients.
The goal is to match each client with at-most-one server. Clients and servers are
abstracted as the vertices of a bipartite graph, and a synchronous, failure-free, message-passing model of computation is assumed, where edges represent communication links.

One can also relate the at-most-once problem to the consensus problem\,\cite{FLP85,MH91,Lynch1996,Paxos98}. 
Indeed, consensus can be viewed as an
at-most-once distributed decision. 
Another related problem is  process renaming, see Attiya \emph{et al.}\,\cite{ABDPR90} where  each process identifier should 
be assigned to at most one process. 

The at-most-once problem has also 
many similarities with the Write-All problem
for the shared memory model\,\cite{AW97,ChlebusKowal05,JHMV01,Alex97,KS08,Malewicz05}.
First presented by Kanellakis and Shvartsman\,\cite{Alex97},
the Write-All problem 
is concerned with performing each job {\em at-least-once}. 
Most of the solutions for the Write-All problem, exhibit
super-linear work even when $m \ll n$. Malewicz\,\cite{Malewicz05}
was the first to present a solution for the Write-All problem
that has linear work for a non-trivial number of processors.
The algorithm presented by Malewicz\,\cite{Malewicz05}
has work $\mathrm{O}(n +  m^{4} \log n)$ and uses test-and-set operations.
Later Kowalski and Shvartsman\,\cite{KS08} presented a solution for 
the Write-All problem that for any constant $\epsilon$ has work 
$\mathrm{O}(n +  m^{2+\epsilon})$. Their algorithm uses a collection
of $q$ permutations with contention $\mathrm{O}(q \log q)$ for a properly chosen constant $q$
 and does not rely on  test-and-set operations. 
Although an efficient polynomial time construction
of permutations with contention $\BigO{q \text{ polylog } q}$ has been 
developed by Kowalski \emph{et al.}\,\cite{Kowalski05}, it is not known to 
date how to construct permutations  with contention $\BigO{q \log q}$ in polynomial time. 
Subsequent to the conference version of this paper\,\cite{KK12}, 
Alistarh \emph{et al.}\,\cite{Dan12} show that there exists a deterministic algorithm 
for the Write-All problem with work $\BigO{n+ m \log^{5} n \log^{2} \max(n,m)}$, by 
derandomizing their randomized solution for the problem. Their solution is a breakthrough in
terms of bridging the gap between the $\Omega\left(n + m \log m\right)$ lower bound 
for the Write-All problem and known deterministic solutions, but is so far existential.
For a detailed overview of research on the Write-All problem, we refer the
reader to the books by Georgiou and Shvartsman\,\cite{GS08, GS11}. 

We note that
the at-most-once problem becomes much simpler
when shared-memory is supplemented by some type of
read-modify-write operations. 
For example, one can associate a \emph{test-and-set} 
bit with each job, ensuring that the job is assigned
to the only process that successfully sets 
the shared bit. An effectiveness optimal implementation
can then be easily obtained from any Write-All solution. 
In this paper we deal only with  the more challenging setting where algorithms
use atomic read/write registers. 

\Par{Contributions:}
We present and analyze the algorithm $\algName$ that solves the at-most-once
problem. The algorithm is parametrized by  $\beta \geq m$ and has effectiveness $n-\beta-m+2$. 
If $\beta < m$ the correctness of the algorithm is still guaranteed, but the termination of the
algorithm cannot be guaranteed. 
For  $\beta = m$ the algorithm 
has optimal effectiveness of $n-2m+2$ up to an additive factor of $m$.
Note that the upper bound for the effectiveness 
of any algorithm  is $n-f$ \cite{KKNS09}, where $f \leq m-1$ is the number of failures in the system. 
We further prove that for  $\beta \geq 3m^2$ the algorithm has work complexity $\BigO{n m \log n \log m}$. 
%
We use algorithm $\algName$ with $\beta = 3m^2$, in order to construct an iterated version of our algorithm 
which for any constant $\epsilon > 0$, has effectiveness of $n-\BigO{m^2 \log n \log m}$ and work 
complexity $\BigO{n + m^{3+\epsilon} \log n}$. This is both effectiveness-optimal and 
work-optimal for any $m=\BigO{\sqrt[3+\epsilon]{n/ \log n}}$.
We note that our solutions are deterministic and assume worst-case behavior. In the probabilistic setting Censor-Hillel\,\cite{Hillel2010} and Kentros\,\emph{et al.}\,\cite{SCA12}
show that optimal effectiveness can be achieved with expected work complexity 
$\BigO{nm^2\log m}$ and $\BigO{n + m^{2+\epsilon}\log n}$, for any small constant $\epsilon$,
respectively.

We then demonstrate how to use the iterated version of our algorithm in order to solve the Write-All problem
with work complexity  $\mathrm{O}(n + m^{3+\epsilon} \log n)$ for any constant $\epsilon > 0$.
Our solution improves on the algorithm of Malewicz\,\cite{Malewicz05}, which 
solves the Write-All problem for a non-trivial number of processes with optimal (linear) work complexity, 
in two ways. First our solution
is work optimal for a wider range of choices for $m$, namely for any $m = \BigO{\sqrt[3+\epsilon]{n / \log n}}$,
cf. the restriction $m = \BigO{\sqrt[4]{n / \log n}}$ of Malewicz, \cite{Malewicz05}. 
Second our solution does not assume the test-and-set primitive used by Malewicz\,
and relies \emph{only} on atomic read/write memory.
There is also a Write-All algorithm due to Kowalski and Shvartsman~\cite{KS08}, which 
does not use test-and-set operations and is work optimal 
for a wider range of processors $m$
than our algorithm, specifically for $m = \BigO{\sqrt[2+\epsilon]{n}}$. However, their algorithm uses a collection
of $q$ permutations with contention $\BigO{q \log q}$ and it is not known to date how to construct 
such permutations in polynomial time (see the discussion in the related work section). Finally, 
subsequent to the conference version of this paper\,\cite{KK12}, 
Alistarh \emph{et al.}\,\cite{Dan12} show that there exists a deterministic algorithm 
for the Write-All problem with work $\BigO{n+ m \log^{5} n \log^{2} \max(n,m)}$. 
Their solution is so far existential, while ours explicit.

\Par{Outline:} 
In Section\,\ref{sec:ModelAll} we formalize the model and introduce definitions
and notations used in the paper. 
In Section\,\ref{sec:Optimal alg} we present the algorithm $\algName$. 
In Sections\,\ref{sec:Effectiveness} and\,\ref{sec:work} we analyze correctness,
effectiveness and work complexity of algorithm $\algName$. In Section\,\ref{sec:ItAlg} 
we present and analyze the iterative algorithm $\ItAlg{\epsilon}$. In Section\,\ref{sec:WAItAlg} 
we present and analyze the iterative algorithm $\WItAlg{\epsilon}$ for the Write-All problem.
Finally, we conclude with Section\,\ref{sec:conclude}.


\section{Model, Definitions, and Efficiency}
\label{sec:ModelAll}
We  define our model,
the at-most-once problem, and measures of efficiency.

\subsection{Model and Adversary}
\label{sec:PRAMmodel}
We model a multi-processor as 
$m$ asynchronous, crash-prone
processes with unique identifiers from some set $\cal P$.
Shared memory is modeled as 
a collection of atomic read/write memory cells,
where the number of bits in each cell is explicitly defined.
%
%
We use the \emph{Input/Output Automata} formalism\,\cite{LT89,Lynch1996}
to specify and reason about algorithms; 
specifically, we use
the \emph{asynchronous shared memory automaton} formalization\,\cite{GL90,Lynch1996}. 
%
Each process $p$ 
is defined in terms of its states
$states_{p}$ and its actions $acts_p$, where
each action is of the type \emph{input}, \emph{output},
or \emph{internal}.
A subset $start_{p}\subseteq states_{p}$ contains 
all the start states of $p$.
%
Each shared variable $x$ takes values 
from a set $V_x$, among which there
is $init_x$, the initial value of $x$. 

We model an algorithm $A$
as a composition
of the automata for each process $p$.
Automaton $A$ consists of a set of states $states(A)$,
where each state $s$ contains a state $s_p\in states_p$ for each $p$, 
and a value $v\in V_x$ for each shared variable $x$. 
Start states $start(A)$ is a subset of $states(A)$, where each state
contains a $start _p$ for each $p$ and an $init_x$ 
for each  $x$. 
%
%
The actions of $A$, $acts(A)$ consists of actions $\pi\in acts_p$
for each process $p$. A transition is the modification of the 
state
as a result of an action and is represented by a triple
($s,\pi,s'$), where $s,s'\in states(A)$ and $\pi\in acts(A)$.
State $s$ is called the \emph{enabling} state of action $\acts$.
The set of all transitions is denoted by $trans(A)$. 
%
Each action in $acts(A)$ is performed by a process, 
thus for any transition ($s,\pi,s'$), $s$ and $s'$ may differ only with respect 
to the state $s_p$ of  process $p$ that invoked $\pi$
and potentially the value of the shared variable that $p$
 interacts with during~$\pi$. 
We also use triples
 $(\{vars_s\},\pi,\{vars_{s'}\})$, where $vars_s$ and $vars_{s'}$ 
are subsets of variables in $s$ and $s'$ respectively, as a shorthand to describe
transitions without having to specify $s$ and $s'$ completely;
here $vars_s$ and $vars_{s'}$ 
 contain only the variables whose value changes as the result
of $\pi$, plus possibly some other variables of interest.
%
%

\ignore{
We say that  states $s$ and $t$ in $states(A)$
are  \emph{indistinguishable} to process $p$ if: 
1)
$s_p=t_p$, and 
2) the values of all shared variables are the same in $s$ and
$t$. 
Now, if states $s$ and $t$ are indistinguishable to
$p$ and 
$(s,\pi,s') \in trans(A)$ for $\pi\in acts_p$, 
then $(t,\pi,t') \in trans(A)$, 
and $s'$ and $t'$ are also indistinguishable to~$p$.
}
\sloppypar{ 
An \emph{execution} fragment of $A$ is either a 
finite sequence, 
$s_{0}$,$\acts_{1}$,$s_{1}$,
$\ldots$,$\acts_{r}$,$s_{r}$, 
or an infinite sequence, 
$s_{0}$,$\acts_{1}$,$s_{1}$,
$\ldots$,$\acts_{r}$,$s_{r}$,$\ldots$, 
of alternating states and actions,
where $(s_{k},\acts_{k+1},s_{k+1}) \in trans(A)$ for 
any
$k\geq0$.
If $s_0 \in start(A)$, then the sequence is called an \emph{execution}. 
The set of executions of $A$ is \emph{execs(A)}. 
We say that  execution $\EX$
is \emph{fair}, 
if $\EX$ is finite and its last state is a state
of $A$ where 
no locally controlled action is enabled, 
or $\EX$ is infinite and every locally controlled action $\pi\in acts(A)$ 
is performed infinitely many times or 
there are infinitely many states in $\EX$ where
$\pi$ is disabled. 
The set of fair executions of $A$ is 
$\fe(A)$.
An execution fragment $\EX'$
\emph{extends} a finite execution fragment $\EX$ of $A$,
if $\EX'$ begins with the last state of $\EX$.
We let $\EX\cdot\EX'$ stand for the execution fragment
resulting from concatenating $\EX$ and $\EX'$ and removing
the (duplicated) first state of $\EX'$.
%
%
}
\ignore{
 A state is said to be \emph{reachable} in $A$ if it 
is the final state of a finite execution of $A$. 
The \emph{trace} of an execution $\EX$ of $A$, denoted by $trace(\EX)$, is a subsequence of $\EX$
consisting of all the external actions, i.e., those actions in  $ ext\_sig(A)$. We say that $\beta$ is a
\emph{trace} of $A$ if $\beta$ is the trace of an execution of $A$. We denote the set of traces of 
$A$ by $traces(A)$.
}

For two states $\st$ and $\st'$ of an execution fragment $\EX$, we say that state $\st$ \emph{precedes} state $\st'$ and we write
$\st < \st'$ if $\st$ appears before $\st'$ in $\EX$. Moreover we write $\st \leq \st'$ if state $\st$ 
either precedes state $\st'$ in $\EX$ or the states $\st$ and $\st'$ are the same state of $\EX$. We use the term precedes and the symbols
$<$ and $\leq$ in a same way for the actions of an execution fragment. We use  the term precedes and the symbol $<$ if an action
$\acts$ appears before a state $\st$ in an execution fragment $\EX$ or if a state $\st$ appears before an action $\acts$ in $\EX$.
Finally for a set of states $S$ of an execution fragment $\EX$, we define as $\st_{max} = \max{S}$ the state $\st_{max} \in S$,
s.t. $\forall \st \in S$,  $\st \leq \st_{max}$ in $\EX$. 
 
We model process crashes by
action $\act{stop}_p$ in $acts(A)$
for each process $p$.
If $\act{stop}_p$ appears in an execution $\EX$
then no actions $\pi\in acts_p$ appear in $\EX$ thereafter.
We then say that process $p$ \emph{crashed}.
%
Actions $\act{stop}_p$ arrive from 
some unspecified external environment, called an \emph{adversary}.
In this work we consider an  \emph{omniscient}, \emph{on-line adversary}\,\cite{Alex97}
that has complete knowledge of the algorithm \sk{executed by the processes}. 
The adversary controls asynchrony and crashes.
We allow up to $f < m$ crashes.
We denote by $\fe_{f}(A)$ all  
fair executions of $A$ with at most $f$ crashes.
\sk{Note that since the processes can only communicate through atomic read/write operations in the
shared memory, all the asynchronous executions are linearizable. This means that concurrent actions 
can be mapped to an equivalent sequence of state transitions, where only
one process performs an action in each transition, and thus the model presented above is appropriate
for the analysis of a multi-process asynchronous atomic read/write shared memory system.}
\BB\subsection{At-Most-Once Problem, Effectiveness and Complexity}
\label{ssec:AMOproblem}

Let $A$ be an algorithm
specified for $m$ processes
with ids from set $\pset=[1\ldots m]$,
and for $n$ jobs with unique ids from set $\tset=[1\ldots n]$.
We assume that there are at least 
as many jobs as there are processes, i.e., $n\ge m$.
We model the performance of job $j$ by process $p$
by means of action $\act{do}_{p,j}$.
For a sequence $c$, we let $len(c)$  denote its length,
and we let $c|_\pi$ denote the sequence of elements $\pi$
occurring in $c$.  
Then for an execution $\EX$, $\doset{\EX}{p}{j}$ is the number
of times process $p$ performs job  $j$. Finally we denote by $F_\EX = \left\{p| stop_p\text{ occurs in }\EX\right\}$ 
the set of crashed processes in
execution $\EX$.
Now we define the number of jobs performed in an execution. Note here that we are borrowing most definitions from Kentros \emph{et al.}\,\cite{KKNS09}. 

\begin{definition}
\label{def:donum}
For  execution $\EX$
let $\tset_\EX = \{j\in\tset|\act{do}_{p,j}$
$\text{occurs in~} \EX \text{~for some~} p\in \pset \}$. 
The total number of jobs performed in $\EX$ is defined to be
	$\donum{\EX}=\left| \tset_{\EX} \right|$.
\end{definition}
\ignore{\text{ for some p} \in \pset}%
We next define the \emph{at-most-once} problem. 
%
%
%
%
\begin{definition}
\label{def:atm1}
Algorithm $A$ solves the at-most-once problem if for each execution $\EX$ of $A$
we have $\forall j\in\tset:\sum_{p\in\pset}\doset{\EX}{p}{j}\leq 1$. 
\end{definition}

\begin{definition}
\label{def:rank}
Let $S$ be a set of elements with unique identifiers. We define as the rank of element $x \in S$
and we write $\rank{x}{S}$, the rank of $x$ if we sort \sk{in ascending order} the elements of $S$ according to 
their identifiers.  \BB   
\end{definition}
%
%
%
%
\subsubsection*{Measures of Efficiency}
We analyze our algorithms in terms of two  
complexity measures: \emph{effectiveness} and \emph{work}.
Effectiveness counts the number of jobs performed by an algorithm in the worst case. 

\begin{definition}
\label{def:eff}
$E_A(n,m,f)=\min_{\EX\in \fe_{f}(A)}(\donum{\EX})$ is the 
\textbf{effectiveness} of algorithm $A$, where $m$ is the number of processes, $n$ is the number of jobs, 
and $f$ is the number of crashes. 
%
%
\end{definition}

A trivial algorithm can solve the at-most-once problem
by splitting the $n$ jobs in groups of size $\frac{n}{m}$ 
and assigning one group to each process.
\nn{Such a solution has effectiveness
$E(n,m,f) = (m-f)\cdot \frac{n}{m}$ (consider an execution where $f$ processes fail 
at the beginning of the execution).} 

\ignore{
Work complexity measures the efficiency of an algorithm in terms 
of the total number of memory 
accesses.
}
Work complexity measures the total number of basic operations
(comparisons, additions, multiplications, shared memory reads and writes) performed by an algorithm. 
We assume that each internal or shared memory cell has size $\mathrm{O}(\log n)$ bits and performing operations involving
a constant number of memory cell costs $\mathrm{O}(1)$. This is consistent with the way work complexity is measured 
in previous related work\,\cite{Alex97,KS08,Malewicz05}.
%
\begin{definition}
\label{def:work}
The {\bf \emph{work}} of algorithm $A$, denoted by $W_A$, 
is the worst case total number of basic operations 
performed by all the processes of algorithm $A$.  
\end{definition}
\ignore{
Space complexity measures the shared memory space used by the algorithm.
\begin{definition}
\label{def:space}
The {\bf \emph{space}} of algorithm $A$
is the 
total number of shared memory cells 
used by the processes of $A$. \BB
\end{definition}
}
\ignore{
Finally here we give definitions of wait-free protocols for the \emph{consensus} problem and the 
\emph{job-assignement} problem. The later is the decision problem related with
the do-exactly-once problem. 
}


Finally we repeat here as a theorem, Corollary\,$1$ from Kentros et al.\,\cite{KKNS09}, that gives an upper bound on the 
  effectiveness for any algorithm solving the  at-most-once problem. \BB

\begin{theorem}{from Kentros et al.\,\cite{KKNS09} \\}
\label{thm:L-Bound}
For all algorithms $A$ that solve the at-most-once problem
with $m$ processes and $n\geq m$ jobs  
in the presence of $f<m$ crashes 
it holds that  $E_A(n,m,f) \leq n-f$. \BB
\end{theorem}
%
\section{Algorithm $\algName$}
\label{sec:Optimal alg}
We present algorithm $\algName$, that solves the at-most-once problem. 
Parameter $\beta \in \mathbb{N}$ is the termination parameter of the 
algorithm. Algorithm $\algName$ is defined for all $\beta \geq m$
\sk{. If $\beta = m$, algorithm $\algName$} 
has optimal up to an additive factor of $m$ effectiveness. 
Note that although 
 $\beta \geq m$ is not necessary in order to prove the correctness of the algorithm, if 
 $\beta < m$ we cannot guarantee termination of algorithm $\algName$. 
\begin{figure*}[!ht]
	\hrule
	\FF
	{\scriptsize \noindent{\bf Shared Variables:}} \\
			{
				\scriptsize
				\T $next=\{next_1,\ldots,next_m\}$, ~$next_q \in \{0,\ldots,n \}$  initially $0$ \\
				\T $done=\{done_{1,1},\ldots,done_{m,n}\}$, ~$done_{q,i} \in \{0,\ldots,n \}$  initially $0$
			}\hfill\vfill
	{\scriptsize \noindent{\bf Signature:}}\\
	{\centering
			\fourcolcode
			{0.24}{0.24}{0.25}{0.22}
			{
				\scriptsize
				\noindent Input:\\
				\T $\act{stop}_{p}$, ~$p\in\pset$ \\
				\T Output:\\
				\T $\act{do}_{p, j}$, ~$p\in\pset$, ~$j\in\tset$ \\
				\hfill\vfill
			}			
			{
				\scriptsize
			  \noindent Internal:\\
  			\T $\act{compNext}_{p}$, ~$p\in\pset$ \\
  			\T $\act{check}_{p}$, ~$p\in\pset$ 
  			\hfill\vfill
			 }
			 {
			 \scriptsize
			 \noindent Internal \emph{Read:} \\ 
  			\T$\act{gatherTry}_{p}$, ~$p\in\pset$ \\
  			\T$\act{gatherDone}_{p}$, ~$p\in\pset$ 
  			\hfill\vfill
  			}
  			{
  			\scriptsize
  			\noindent Internal \emph{Write:} \\ 
  			\T$\act{setNext}_{p}$, ~$p\in\pset$ \\
  			\T$\act{done}_{p}$, ~$p\in\pset$ \\
  			\hfill\vfill
			 }
	}

	{\scriptsize \noindent{\bf State:}}\\
			{\scriptsize
				\T $\status{p} \in \left\{comp\_next, set\_next, gather\_try, gather\_done, check,do, done, end, stop\right\}$,\\ 
				\T initially $\status{p} = comp\_next$ \\
				\T $\FREE_{p}, \DONE_{p}, \TRY_{p} \subseteq \tset$, ~initially $\FREE_{p}=\tset$ and $\DONE_{p}=\TRY_{p}=\emptyset$\\
				\T $\posA{p}=\left\{\pos{p}{1},\ldots,\pos{p}{m}\right\}$, where $\pos{p}{i} \in \left\{1,\ldots,n\right\}$, initially $\pos{p}{i} = 1$\\	
			}
			\twocolcode{0.54}{0.54}
			{\scriptsize\hspace*{-3mm}
				\T ~ $\nxt{p} \in \left\{1,\ldots,n\right\}$, initially undefined\\
				\T $\tmp{p} \in \left\{0,\ldots,n\right\}$,~initially undefined\\
				\hfill\vfill\vspace*{-2mm}
			}
			{\scriptsize\hspace*{-3mm}
				\T $\q{p} \in \left\{1,\ldots,m\right\}$,~initially 1\\ \\
				\hfill\vfill
			}
	\hrule
		\caption{Algorithm $\algName$: Shared Variables, Signature and States}
		\BB
		\label{fig:IOAsig}
\end{figure*}			

The idea behind the algorithm $\algName$
(see Fig.
\,\ref{fig:IOAsig},\,\ref{fig:IOAtrans}) is quite intuitive and is based on an algorithm
for renaming processes presented by Attiya\,\emph{et al.}\,\cite{ABDPR90}. 
Each process $p$,
picks a job $i$ to perform, announces (by writing in shared memory) that it is about to perform 
the job and then checks if it is safe to perform it (by reading the announcements other processes 
made in the shared memory, and the jobs other processes announced they have performed). If it is 
safe to perform the job $i$, process $p$ will proceed with the 
$\act{do}_{p,i}$ action and then mark the job completed. If it is not safe to perform $i$, $p$
will release the job. In either case, $p$ picks a new job to perform.
In order to pick a new job, $p$ reads from the shared memory and 
gathers information on which jobs 
\sk{are safe to perform, by reading the announcements that other processes 
made in the shared memory about the jobs they are about to perform, 
and the jobs other processes announced they have already performed.} 
Assuming that those jobs are ordered, $p$ splits the set of ``free'' 
jobs in $m$ intervals and picks the first job of the interval with
rank equal to $p$'s rank.
\sk{Note that since the information needed in order to decide whether it is safe to perform a specific job 
and in order to pick the next job to perform is the same, these steps are combined in the algorithm.} 
In Figure \ref{fig:IOAtrans}, we use function $rank(\SET_1, \SET_2, i)$, that returns the element
of set $\SET_1 \setminus \SET_2$ that has rank $i$. If $\SET_1$ and $\SET_2$ have $\BigO{n}$ elements and
are stored in some tree structure like \emph{red-black tree} or some variant of \emph{B-tree}, the
operation $rank(\SET_1, \SET_2, i)$, costs $\BigO{\left|\SET_2\right| \log n}$ assuming that 
$\SET_2 \subseteq \SET_1$.

We will prove that algorithm $\algName$ has effectiveness $n-(\beta+m-2)$. For $\beta = O(m)$ this effectiveness is asymptotically optimal for any $m=o(n)$. Note that by Theorem\,\ref{thm:L-Bound} the upper bound on effectiveness of the at-most-once problem is $n-f$,
where $f$ is the number of failed processes in the system. 
%
%
Next we present algorithm $\algName$ in more detail.

\Par{Shared Variables.}
$next$ is an array with $m$ elements. In the cell $next_q$ of the array process $q$ announces the job it is about to perform. From the structure of algorithm 
$\algName$, only process $q$ writes in cell $next_q$. On the other hand any process
may read cell $next_q$.

$done$ is an $m~\times~n$ matrix. In line $q$ of the matrix, process $q$ announces the
jobs it has performed. Each cell of line $q$ contains the identifier of exactly one
job that has been performed by process $q$. Only process $q$ writes in the cells of
line $q$ but any process may read them. Moreover, process $q$ updates line $q$ by
adding entries at the end of it.
\begin{figure*}[!ht]
	\BB 
	\hrule
	\FF
	{\scriptsize \noindent{\bf Transitions of process $p$:}}
	\begin{center}
		\BBB\BB
			\twocolcode{0.49}{0.49}
		{\scriptsize
			\ef{$\act{{\rm \bf Input~} stop}_{p}$}
			{
		 	$\status{p} \gets stop$
			}
			\BB \BB
		  \prcef{$\act{{\rm \bf Internal~} compNext}_{p}$}
			{
			 $\status{p} = comp\_next$
			}
			{
			{\bf if} $\left|\FREE_{p}\setminus \TRY_p\right| \geq \beta$ {\bf then}\\						
			~ $\tmp{p} \gets \frac{\left|\FREE_p\right|-\left(m-1\right)}{m}$\\
			~ {\bf if} $\tmp{p} \geq 1$ {\bf then}\\
			~ ~ $\tmp{p} \gets \left\lfloor \left(p-1\right)\cdot \tmp{p}\right\rfloor +1$\\ 
			~ ~ $\nxt{p} \gets rank\left(\FREE_p, \TRY_p, \tmp{p} \right)$\\
			~ {\bf else} \\
			~ ~ $\nxt{p} \gets rank\left(\FREE_p, \TRY_p, p \right)$\\
			~ {\bf end} \\
			~ $\q{p} \gets 1$\\
			~ $\TRY_p \gets \emptyset$\\
			~ $\status{p} \gets set\_next$\\
			{\bf else} \\
			~ $\status{p} \gets end$ \\
			{\bf end}
			}
			\BB \BB
			\prcef{$\act{{\rm \bf Internal~Write~} setNext}_{p}$}
    	{
    		$\status{p} = set\_next$
    	}
    	{
      	$ next_p \gets \nxt{p}$\\
      	$ \status{p} \gets gather\_try$
      }
			\prcef{$\act{{\rm \bf Internal~Read~} gatherTry}_{p}$}
			{
			 $\status{p} = gather\_try$
			}
			{
				  {\bf if} $\q{p} \neq p$ 
					~ {\bf then} \\
					~	$\tmp{p} \gets next_{\q{p}}$\\
					~ {\bf if} $\tmp{p} > 0$ 
					~ {\bf then}\\
					~ ~ $\TRY_{p} \gets \TRY_{p} \cup \left\{\tmp{p}\right\} $\\
					~ {\bf end}\\ 
					{\bf end}\\
					{\bf if} $\q{p}+1 \leq m$ 
					{\bf then}\\ 
					~ $\q{p} \gets \q{p}+1$\\
					{\bf else} \\
					~ $\q{p} \gets 1$\\
					~ $\status{p} \gets gather\_done$\\
					{\bf end}
			}
			
		}
			{\scriptsize
			\prcef{$\act{{\rm \bf Internal~Read~} gatherDone}_{p}$}
			{
			 $\status{p} = gather\_done$
			}
			{
					{\bf if} $\q{p} \neq p$ {\bf then} \\
					~	$\tmp{p} \gets done_{\q{p},\pos{p}{\q{p}}}$\\
					~ {\bf if} $\pos{p}{\q{p}} \leq n$ 
					~ AND $\tmp{p} > 0$\\ 
					~ {\bf then}\\
					~ ~ $\DONE_{p} \gets \DONE_{p} \cup \left\{\tmp{p}\right\} $\\
					~ ~ $\FREE_{p} \gets \FREE_{p} \setminus \left\{\tmp{p}\right\} $\\					
					~ ~ $\pos{p}{\q{p}} = \pos{p}{\q{p}} + 1$ \\
					~ {\bf else} $\q{p} \gets \q{p}+1$\\
					~ {\bf end}\\
					{\bf else} $\q{p} \gets \q{p}+1$\\
					{\bf end}\\
					{\bf if} $\q{p} > m$ {\bf then} \\
					~ $\q{p} \gets 1$\\
					~ $\status{p} \gets check$\\
					{\bf end}	
			}	
			\prcef{$\act{{\rm \bf Internal~} check}_{p}$}
    	{
    		$\status{p} = check$
    	}
    	{
      	{\bf if} $\nxt{p} \notin \TRY_p$ AND $\nxt{p} \notin \DONE_p$\\
      	{\bf then}
      	 	$\status{p} \gets do$ \\
      	{\bf else}\\
      	~	$\status{p} \gets comp\_next$\\
      	{\bf end}	
      }
			\BB \BB
			\prcef{$\act{{\rm \bf Output~} do}_{p,j}$}
			{
			 $\status{p} = do$\\
			 $\nxt{p} = j$
			}
			{
				$\status{p} \gets done$
			}
			\BB \BB
			\prcef{$\act{{\rm \bf Internal~Write~} done}_{p}$}
			{
			 $\status{p} = done$
			}
			{
				$done_{p,\pos{p}{p}} \gets \nxt{p}$ \\
				$\DONE_p \gets \DONE_p \cup \left\{\nxt{p}\right\}$ \\
				$\FREE_p \gets \FREE_p \setminus \left\{\nxt{p}\right\}$ \\
				$\pos{p}{p} \gets \pos{p}{p}+1$ \\
				$\status{p} \gets comp\_next$
			}
		}
		\end{center}
	\vspace*{-1mm}\hrule
		\caption{Algorithm $\algName$: Transitions}
		\BB 
		\label{fig:IOAtrans}
\end{figure*}

\Par{Internal Variables of process $p$.}
The variable 
$\status{p}$ 
records the status of process $p$ and defines its next action as follows: $\status{p}=comp\_next$ - process $p$ is ready to compute the next job to perform (this is the initial status of $p$), $\status{p}=set\_next$ - $p$ computed the next job to perform and is ready to announce it by writing in the shared memory, 
$\status{p}=gather\_try$ - $p$ reads the array 
$next$ in shared memory in order to compute the $\TRY_{p}$ set,  $\status{p}=gather\_done$ - $p$ reads the matrix $done$ in shared memory in order to update the 
$\DONE_{p}$ and $\FREE_{p}$ sets, $\status{p}=check$ - $p$ has to check whether it is safe to perform
its current job, $\status{p}=do$ - $p$ can safely perform its current job, $\status{p}=done$ - $p$ performed 
its current job and needs to update the shared memory, $\status{p}=end$ - $p$ terminated, $\status{p}=stop$ - $p$ crashed.
%
\ignore{
$\FREE_{p}, \DONE_{p}, \TRY_{p} \subseteq \tset$ are three sets that are used by process $p$
in order to compute the next job to perform and whether it is safe to perform it. $\FREE_p$, is initially set to $\tset$ and contains an estimate of the jobs
that are still available. $\DONE_{p}$ is initially empty and contains an estimate of the
jobs that have been performed.  No job is removed from $\DONE_{p}$ during the execution of algorithm $\algName$. $\TRY_{p}$ is initially empty and
contains an estimate of the jobs that other processes are about to perform.
}

$\FREE_{p}, \DONE_{p}, \TRY_{p} \subseteq \tset$ are three sets that are used by process $p$
in order to compute the next job to perform and whether it is safe to perform it. We use some tree structure like
\emph{red-black tree} or some variant of \emph{B-tree}\,\cite{Bayer72,GuibasS78} 
for the sets $\FREE_{p}$, $\DONE_{p}$ and $\TRY_{p}$, in order
to be able to add, remove and search elements in them with $\BigO{\log n}$ work. 
$\FREE_p$, is initially set to $\tset$ and contains an estimate of the jobs
that are still available. $\DONE_{p}$ is initially empty and contains an estimate of the
jobs that have been performed.  No job is removed from $\DONE_{p}$ or added to $\FREE_p$ during the execution of algorithm $\algName$. 
$\TRY_{p}$ is initially empty and contains an estimate of the jobs that other processes are about to perform. It holds that
$\left|\TRY_{p}\right| < m$, since there are $m-1$ processes apart from process $p$ that may be attempting to perform a job.

$\posA{p}$ is an array of $m$ elements. Position $\pos{p}{q}$ of the array contains a pointer in the line $q$ of the shared matrix $done$. $\pos{p}{q}$ is the element
of line $q$ that process $p$ will read from. In the special case where $q=p$, $\pos{p}{p}$ is the element of line $p$ that process $p$ will write into after performing a new job. The elements of the shared matrix $done$ are read when process
$p$ is updating the $\DONE_p$ set.

$\nxt{p}$ contains the job process $p$ is attempting to perform.

$\tmp{p}$ is a temporary storage for values read from the shared memory.

$\q{p} \in \left\{1,\ldots,m\right\}$ is used as indexing for looping through process
identifiers.


\Par{Actions of process $p$.} We visit them one by one below. 
%
\ignore{ 
$\act{compNext}_p$: Process $p$ computes the $\FREE_p$ set from the sets $\DONE_p$ and $\TRY_p$.
If the $\FREE_p$ set has more or equal elements to $\beta$, where $\beta$ is the termination parameter of the
algorithm, process $p$ computes its next candidate job, by splitting the $\FREE_p$ set in 
$m$ parts and picking the first element of the $p$-th part. In order to do that it uses the 
function $rank(\SET,i)$, which returns the element of set $\SET$ with rank $i$. Finally process
$p$ sets the $\TRY_p$ set to the empty set, the $\q{p}$ internal variable to 1 and its status 
to $set\_next$ in order to update the shared memory with its new candidate job. If the $\FREE_p$ 
set has less than $\beta$ elements process $p$ terminates.
}

$\act{compNext}_p$: Process $p$ computes the set $\FREE_p \setminus \TRY_p$ and if it has more or equal elements to $\beta$, 
were $\beta$ is the termination parameter of the
algorithm, process $p$ computes its next candidate job, by splitting the $\FREE_p \setminus \TRY_p$ set in 
$m$ parts and picking the first element of the $p$-th part. In order to do that it uses the 
function $rank(\SET_1, \SET_2, i)$, which returns the element of set $\SET_1 \setminus \SET_2$ with rank $i$. 
Finally process $p$ sets the $\TRY_p$ set to the empty set, the $\q{p}$ internal variable to 1 and its status 
to $set\_next$ in order to update the shared memory with its new candidate job. If the $\FREE_p \setminus \TRY_p$ 
set has less than $\beta$ elements process $p$ terminates.

$\act{setNext}_p$: Process $p$ announces its new candidate job by writing the contents of its 
$\nxt{p}$ internal variable in the $p$-th position of the $next$ array. Remember that the $next$
array is stored in shared memory. Process $p$ changes its status to $gather\_try$, in order to
start collecting the $\TRY_p$ set from the $next$ array.

$\act{gatherTry}_p$: With this action process $p$ implements a loop, which reads from the shared memory
all the positions of the array $next$ and updates the $\TRY_p$ set. In each execution of the action, 
process $p$ checks if $\q{p}$ is equal to $p$. If it is not equal, $p$ reads the $\q{p}$-th position of 
the array $next$, checks if the value read is greater than $0$ and if it is, adds the value it read in 
the $\TRY_p$ set. If $\q{p}$ is equal to $p$, $p$ just skips the step described above.
Then $p$ checks if the value of $\q{p}+1$ is less than $m+1$. If it is, then $p$ increases $\q{p}$ by 1
and leaves its status $gather\_try$, otherwise $p$ has finished updating the $\TRY_p$ set and thus sets 
$\q{p}$ to 1 and changes its status to $gather\_done$, in order to update the $\DONE_p$ and 
$\FREE_p$ sets from the contents of the $done$ matrix.

$\act{gatherDone}_p$: With this action process $p$ implements a loop, which updates the $\DONE_p$ and $\FREE_p$ sets with
values read from the matrix $done$, which is stored in shared memory. In each execution of the action, 
process $p$ checks if $\q{p}$ is equal to $p$. If it is not equal, $p$ uses the internal variable 
$\pos{p}{\q{p}}$, in order to read fresh values from the line $\q{p}$ of the $done$ matrix. In detail,
$p$ reads the shared variable $done_{\q{p},\pos{p}{\q{p}}}$, checks if  $\pos{p}{\q{p}}$ is less than $n+1$ and
if the value read is greater than $0$. If both conditions hold, $p$ adds the value read at the $\DONE_p$ set,
removes the value read from the $\FREE_p$ set and increases $\pos{p}{\q{p}}$ by one. Otherwise, it means that 
either process $\q{p}$ has terminated 
(by performing all the $n$ jobs) or the line $\q{p}$ does not contain any new completed jobs. In either
case $p$ increases the value of $\q{p}$ by 1. The value of $\q{p}$ is increased by 1 also if $\q{p}$
was equal to $p$. Finally $p$ checks whether $\q{p}$ is greater than $m$; if it is, $p$ has completed the loop and
thus changes its status to $check$.

$\act{check}_p$: Process $p$ checks if it is safe to perform its current job. This is done by checking if $\nxt{p}$
belongs to the set $\TRY_p$ or to the set $\DONE_p$. If it does not, then it is safe to perform the job $\nxt{p}$
and $p$ changes its status to $do$. Otherwise it is not safe, and thus $p$ changes its status to $comp\_next$, in order
to find a new job that may be safe to perform.

$\act{do}_{p,j}$: Process $p$ performs job $j$. Note that $\nxt{p}=j$ is part of the preconditions for the action to be enabled in
a state. Then  $p$  changes its status to $done$.

$\act{done}_p$: Process $p$ writes in the $done_{p,\pos{p}{p}}$ position of the shared memory the value of
$\nxt{p}$, letting other processes know that it performed job $\nxt{p}$. Also $p$ adds $\nxt{p}$ to its 
$\DONE_p$ set, removes $\nxt{p}$ from its 
$\FREE_p$ set, increases $\pos{p}{p}$ by 1 and changes its status to $comp\_next$.

$\act{stop}_p$: Process $p$ crashes by setting its status to $stop$. 
\section{Correctness and Effectiveness Analysis}
\label{sec:Effectiveness}
We begin the analysis of algorithm $\algName$, by showing in Lemma\,\ref{lem:simpleCorrect} that $\algName$ solves the at-most-once problem. That is, 
there exists no execution of $\algName$ in which 2 distinct actions $\act{do}_{p,i}$ and
$\act{do}_{q,i}$ appear for some $i \in \tset$ and $p,q \in \pset$. 
We continue the analysis by showing in Theorem\,\ref{thm:effectiveness} that algorithm $\algName$ has effectiveness $E_{\algName}(n,m,f)=n-\left(\beta + m - 2\right)$.
This is done in two steps. First in Lemma\,\ref{lem:noTermination}, we show that
 algorithm $\algName$ cannot terminate its execution if less than 
$n-\left(\beta + m - 1\right)$ jobs are performed. The effectiveness analysis is completed by showing in Lemma\,\ref{lem:simpleEFF}, that the algorithm is wait-free 
(it has no infinite fair executions). In Theorem\,\ref{thm:effectiveness} we combine
the two lemmas in order to show that the effectiveness of algorithm $\algName$ 
is greater that or equal to $n-\left(\beta + m - 2\right)$. Moreover, we show the 
existence of an adversarial strategy, that results in a terminating execution where
 $n-\left(\beta + m - 2\right)$ jobs are completed, showing that the bound is tight.

\sk{In the analysis that follows,} for a state $\st$ and a process $p$ we denote by 
$\FREEinS{\st}{p},~\DONEinS{\st}{p},~\TRYinS{\st}{p}$,
the values of the internal variables $\FREE$, $\DONE$ and $\TRY$ of process $p$ in state $\st$.
Moreover with $\st.next$, and $\st.done$ we denote the contents of the array $next$ and the matrix $done$ in state $\st$. 
Remember that  $next$ and $done$, are stored in shared memory. 
\begin{lemma}
\label{lem:simpleCorrect}
There exists no execution $\EX$ of algorithm $\algName$, such that $\exists i \in \tset$
and $\exists p,q \in \pset$ for which $\act{do}_{p,i}, \act{do}_{q,i} \in \EX$.
\end{lemma}
\begin{proof} 
Let us for the sake of contradiction assume that there exists \sk{an} execution $\EX \in execs(\algName)$ and 
$i \in \tset$ and $p,q \in \pset$ such that $\act{do}_{p,i}, \act{do}_{q,i} \in \EX$. We examine two cases.

\Par{Case 1 $p=q$:}
Let states $\st_1, \st_{1}^{'}, \st_2, \st_{2}^{'} \in \EX$, such that the transitions 
$\left(\st_1, \act{do}_{p,i},\st_{1}^{'}\right)$, $\left(\st_2, \act{do}_{p,i},\st_{2}^{'}\right) \in \EX$ and without loss of generality assume $\st_{1}^{'} \leq \st_2$ in $\EX$. From Figure \ref{fig:IOAtrans} we
have that $\st_{1}^{'}.\nxt{p}=i$, $\st_{1}^{'}.\status{p}=done$ and  $\st_{2}.\nxt{p}=i$, $\st_{2}.\status{p}=do$. 
From algorithm $\algName$,  state $\st_{2}$ must be preceded by transition 
$\left(\st_{3} , \act{check}_{p},\st_{3}^{'}\right)$, such that $\st_{3}.\nxt{p}=i$ and 
$\st_{3}^{'}.\nxt{p}=i$, $\st_{3}^{'}.\status{p}=do$, where $\st_{1}^{'}$ precedes $\st_3$ in $\EX$.
Finally $\st_3$ must be preceded in $\EX$ by transition $\left(\st_{4} , \act{done}_{p},\st_{4}^{'}\right)$,
where $\st_{1}^{'}$ precedes $\st_4$, such that $\st_{4}.\nxt{p}=i$ and $i \in \st_{4}^{'}.\DONE_p$.
Since $\st_{4}^{'}$ precedes $\st_3$ and during the execution of $\algName$ no elements are removed from 
$\DONE_p$, we have that $i \in \st_{3}.\DONE_p$. This is a contradiction, since the transition  
$(\left\{\nxt{p}=i, i\in \DONE_p \right\},$ $\act{check}_{p},$ $\left\{\nxt{p}=i, \status{p}=do\right\})$ $\notin trans(\algName)$.

\Par{Case 2 $p\neq q$:}
Given transition $\left(\st_{1},\act{do}_{p,i},\st_{1}^{'}\right)$ 
in execution $\EX$, we deduce 
from Fig.\,\ref{fig:IOAtrans} that there exist in $\EX$ transitions $\left(\st_2,\act{setNext}_{p},\st_{2}^{'}\right)$,
$\left(\st_3,\act{gatherTry}_{p},\st_{3}^{'}\right)$, $\left(\st_4,\act{check}_{p},\st_{4}^{'}\right)$,
where $\st_{2}^{'}.next_p = \st_{2}^{'}.\nxt{p} = i$, $\st_3.next_p=\st_{3}.\nxt{p}=i,\st_3.\q{p}=q$, $\st_4.\nxt{p}=i$,
$\st_{4}^{'}.\nxt{p}=i$, $\st_{4}^{'}.\status{p}=do$,
such that 
$\st_{2} < \st_{3} < \st_{4} < \st_1$ and
there exists no action $\acts = \act{compNext}_p$ in execution 
$\EX$, such that  $\st_2 < \acts < \st_{1}^{'}$.
%

Similarly for transition $\left(\stt_1,\act{do}_{q,i},\stt_{1}^{'}\right)$
there exist in execution $\EX$ transitions 
$\left(\stt_2,\act{setNext}_{q},\stt_{2}^{'}\right)$,
$\left(\stt_3,\act{gatherTry}_{q},\stt_{3}^{'}\right)$, $\left(\stt_4,\act{check}_{q},\stt_{4}^{'}\right)$,
where
$\stt_{2}^{'}.next_q = \stt_{2}^{'}.\nxt{q} = i$, $\stt_3.next_q=\stt_{3}.\nxt{q}=i,\stt_3.\q{q}=p$,
$\stt_4.\nxt{q}=i$,
$\stt_{4}^{'}.\nxt{q}=i$, $\stt_{4}^{'}.\status{q}=do$,
such that 
 $\stt_{2} < \stt_{3} < \stt_{4} < \stt_{1}$ and
 there exists no action $\acts'=\act{compNext}_q$ in execution 
$\EX$, such that  $\stt_2 < \acts < \stt_{1}^{'}$.

Either state $\st_{2} < \stt_{3}$ or 
$ \stt_{3} < \st_{2}$ which implies  $\stt_{2} < \st_{3}$.
We will show that if $\st_{2} < \stt_{3}$ then $\act{do}_{q,i}$ cannot take place, leading to a contradiction. 
The case where $\stt_{2} < \st_{3}$ is symmetric and will be omitted.

Let us assume that $\st_{2}$ precedes $\stt_{3}$. 
We have two cases, either $\stt_{3}.next_p=i$ or $\stt_{3}.next_p \neq i$. In the first case $i \in \stt_{3}^{'}.\TRY_q$.
The only action in which entries are removed  from the $\TRY_q$ set, is  action $\act{compNext}_q$, 
where the $\TRY_q$ set is reset to $\emptyset$. Thus $i \in \stt_{4}.\TRY_q$, 
since $\nexists ~ \acts' = \act{compNext}_q \in \EX$, such that $\stt_2<\acts'<\stt_1$. This is a contradiction since $\left(\stt_4,\act{check}_{q},\stt_{4}^{'}\right)
\notin trans(\algName)$, if  $i \in \stt_{4}.\TRY_q$, $\stt_{4}.\nxt{q}=i$ and $\stt_{4}^{'}.\status{q}=do$.

If $\stt_{3}.next_p \neq i$, since  $\left(\st_2,\act{setNext}_{p},\st_{2}^{'}\right) \in \EX$ and
$\st_{2}^{'} < \stt_{3}$ there exists action $\acts_1 = \act{setNext}_{p} \in \EX$,
such that $\st_{2}^{'}< \acts_1 < \stt_{3}$. Moreover, there exists action
$\acts_2 = \act{compNext}_p$ in $\EX$, such that $\st_{2}^{'} < \acts_2  < \acts_1$. Since $\nexists ~ \acts = \act{compNext}_p \in \EX$, 
such that $\st_2<\acts <\st_{1}^{'}$,  it holds that $\st_{1}^{'} < \acts_2  < \acts_1 < \stt_{3}$. 
Furthermore, from Fig.\,\ref{fig:IOAtrans}
there exists transition $\left(\st_5,\act{done}_p,\st_{5}^{'}\right)$ in $\EX$ and $j \in \{1,\ldots,n\}$, such that  
$\st_{5}.\pos{p}{p}=j$, $\st_{5}.done_{p,j}=0$, $\st_{5}.\nxt{p}=i$, $\st_{5}^{'}.done_{p,j}=i$ and 
$\st_{1}^{'} < \st_{5}^{'} < \acts_2 < \stt_{3}$. It must be the case that $i \notin \stt_{2}.DONE_q$, since
$\stt_{2}.\nxt{q} = i$. From that and from Fig.\,\ref{fig:IOAtrans} we have that there exists
transition $\left(\stt_6,\act{gatherDone}_q,\stt_{6}^{'}\right)$ in $\EX$, such that $\stt_6.\q{q}=p$,
$\stt_6.\pos{q}{p}=j$ and $\stt_{3} < \stt_6 < \stt_4$. Since $\st_{5}^{'} < \stt_{3}$
and $done_{p,j}$ from  algorithm $\algName$ cannot be changed again in execution $\EX$, we have that 
$\stt_6.done_{p,j}=i$ and as a result $i \in \stt_{6}^{'}.\DONE_q$. Moreover, during the execution of 
algorithm $\algName$, entries in set $\DONE_q$ are only added and never removed, thus we have that $i \in \stt_4.DONE_q$.
This is a contradiction since $\left(\stt_4,\act{check}_{q},\stt_{4}^{'}\right)
\notin trans(\algName)$, if  $i \in \stt_{4}.\DONE_q$, $\stt_{4}.\nxt{q}=i$ and $\stt_{4}^{'}.\status{q}=do$.
This completes the proof.

\end{proof}

Next we examine the effectiveness of the algorithm. First we show that
 algorithm $\algName$ cannot terminate its execution if less than 
$n-\left(\beta + m - 1\right)$ jobs are performed.


\begin{lemma}
\label{lem:noTermination}
For any $\beta \geq m$, $f \leq m-1$ and for any finite execution $\EX \in \ex{\algName}$ 
with $\donum{\EX} \leq n-\left(\beta + m - 1\right)$, there exists \sk{a} (non-empty) execution fragment $\EX'$ such 
that $\EX\cdot\EX' \in \ex{\algName}$.
\end{lemma}
\begin{proof} 
From the algorithm $\algName$, we have that for any process $p$ and any state $\st \in \EX$, 
$\left|\FREEinS{\st}{p}\right| \geq n - \donum{\EX}$ and $\left|\TRYinS{\st}{p}\right| \leq m-1$. 
The first inequality holds since the $\FREEinS{\st}{p}$ set is estimated by $p$ by examining 
the $done$ matrix which is stored in shared memory. 
From algorithm $\algName$, a job $j$ is only inserted in
line $q$ of the matrix $done$, if a $\act{do}_{q,j}$ action has already been performed by process $q$. The second inequality
is obvious.
Thus we have that $\forall p\in \pset$ and 
$\forall \st \in \EX$,
 $\left|\FREEinS{\st}{p} \setminus \TRYinS{\st}{p} \right| \geq n - \left(\donum{\EX}+ m-1\right)$. If $\donum{\EX} \leq n-\left(\beta + m - 1\right)$,
$\forall p\in \pset$ and $\forall \st \in \EX$ we have that $\left|\FREEinS{\st}{p} \setminus \TRYinS{\st}{p} \right| \geq \beta$. 
Since there can be $f\leq m-1$ failed processes in our system, at the final state $\st'$ of execution $\EX$ there exists at least one
process $p \in \pset$ that has not failed. This process has not terminated, since from Fig.\,\ref{fig:IOAtrans} a process $p$
can only terminate if \sk{in} the enabling state $\st$ of action $\act{compNext}_p$, $\left|\FREEinS{\st}{p} \setminus \TRYinS{\st}{p} \right|<\beta$. 
This process can continue executing steps and thus there exists a (non-empty) execution fragment $\EX'$ such that $\EX\cdot\EX' \in \ex{\algName}$.

\end{proof}

Since no finite execution of algorithm $\algName$ can terminate if less than $n-(\beta + m-1)$ jobs are performed, Lemma\,\ref{lem:noTermination} implies that if the algorithm $\algName$ has effectiveness less than or equal to $n-(\beta + m-1)$, there must exist some infinite fair execution $\EX$  with 
$\donum{\EX} \leq n-(\beta + m-1)$. Next we prove 
that  algorithm $\algName$ is wait-free (it has no infinite fair executions).
%

\begin{lemma}
\label{lem:simpleEFF}
For any $\beta \geq m$, $f \leq m-1$  there exists no infinite fair execution $\EX \in execs(\algName)$. 
\end{lemma}
\begin{proof} 
We will prove this by contradiction. 
Let  $\beta \geq m$  and $\EX \in execs(\algName)$ an
infinite fair execution with  $f \leq m-1$ failures, 
and let $\donum{\EX}$ be the jobs executed by execution $\EX$ according to Definition\,\ref{def:donum}.
Since $\EX\in execs(\algName)$ and from Lemma\,\ref{lem:simpleCorrect} $\algName$ solves the at-most-once
problem, $\donum{\EX}$ is finite. Clearly there exists at least one process in 
execution $\EX$ that has
not crashed and does not terminate\,(some process must take steps in $\EX$ in order for it to be infinite).
Since $\donum{\EX}$ and $f$ are finite, there exists \sk{a} state $\st_0$ in $\EX$ 
such that after $\st_0$ no process crashes, no process terminates, no $\act{do}$ action takes place in $\EX$ and no process
adds new entries in the $done$ matrix in shared memory. The later holds since the execution is infinite and fair, 
the $\donum{\EX}$ is also finite, consequently any non failed process $q$ that has not terminated will eventually update the $q$ line of
the $done$ matrix to be in agreement with the $\act{do}_{q,*}$ actions it has performed. Moreover any process $q$ that
has terminated, has already updated the  $q$ line of $done$ matrix with the latest $\act{do}$ action it performed, before it 
terminated, since in order to terminate it must have reached a $\act{compNext}$ action that has set its status to $end$.


We define the following sets of processes and jobs according to state $\st_0$.
$\tset_\EX$ are jobs that have been performed in $\EX$ according to Definition \ref{def:donum}.
$\pset_\EX$ are processes that do not crash and do not terminate in $\EX$. By the way we defined 
state  $\st_0$ only processes in $\pset_\EX$ take steps in $\EX$ after state $\st_0$.
$\Stuck_\EX=\left\{i \in \tset \setminus \tset_\EX| \exists\text{ failed process }p:\st_0.next_p=i\right\}$, i.e., 
$\Stuck_\EX$ expresses the set of 
jobs that are held  by failed processes.
$\DONE_\EX = \left\{i \in \tset_\EX| \exists p \in \pset~\text{and}~j\in\{1,\ldots,n\}:\st_0.done_p(j)=i\right\}$, i.e., 
$\DONE_\EX$ expresses the set of jobs that have been performed
before state $s_0$ and the processes that performed them managed to update the shared memory. 
Finally we define $\Pool_\EX = \tset \setminus (\tset_\EX  \cup \Stuck_\EX)$. 
After state  $\st_0$, all processes in $\pset_\EX$ will keep executing. This means that
whenever a process $p \in \pset_\EX$ takes action $\act{compNext}_p$ in $\EX$, the first if statement is true.
Specifically it holds that for $\forall p \in \pset_\EX$ and for all the enabling states $\st \geq \st_{0}$ of actions 
$\act{compNext}_p$ in $\EX$, $\left|\FREE_p \setminus \TRY_p \right| \geq \beta$.


From Figure \ref{fig:IOAtrans}, we have that for any $p \in \pset_\EX$, 
$\exists~\st_p \in \EX$ 
such that $\st_p > \st_0$ and  for all states $\st \geq \st_{p}$, $\DONEinS{\st}{p}=\DONE_\EX,~
\FREEinS{\st}{p}=\tset \setminus\DONE_\EX ~\text{and}~
\FREEinS{\st}{p} \setminus \TRYinS{\st}{p} \subseteq \Pool_\EX$. Let $\st_{0}'=\max_{p\in \pset_{\EX}}[\st_p]$. 
From the above we have: 
$\left| \tset \setminus\DONE_\EX \right| \geq \beta \geq m$ and
$\left| \Pool_\EX \right|   \geq \beta \geq m$, 
since
$\forall \st' \geq \st_{0}'$ we have that 
$\FREEinS{\st'}{p}=\tset \setminus\DONE_\EX$ and $\FREEinS{\st'}{p} \setminus \TRYinS{\st'}{p} \subseteq \Pool_\EX$ and $\forall p \in \pset_\EX$ 
and for all the enabling states $\st \geq \st_{0}'$ of actions $\act{compNext}_p$ in $\EX$, we have that 
$\left|\FREE_p \setminus \TRY_p \right| \geq \beta$.

Let $p_0$ be the process with the smallest process identifier in $\pset_\EX$. 
We examine $2$ cases according to the size of $\tset \setminus\DONE_\EX$.

{\bf Case A $\left| \tset \setminus\DONE_\EX \right| \geq 2m-1$}: Let $x_{0} \in \Pool_\EX$ be the job such that $\rank{x_{0}}{\Pool_\EX}=\left\lfloor \left(p_{0}-1\right)\cdot \frac{|\tset \setminus \DONE_\EX|-\left(m-1\right)}{m} \right\rfloor+1$.
Such $x_{0}$ exists since $\forall p \in \pset_\EX$ and  $\forall \st \geq \st_{0}'$ it holds 
$\FREEinS{\st}{p} \setminus \TRYinS{\st}{p} \subseteq \Pool_\EX$, $\FREEinS{\st}{p} = \tset \setminus \DONE_\EX$
from which we have that $\left|\Pool_\EX\right| \geq \left|\tset \setminus \DONE_\EX\right| - \left|\TRYinS{\st}{p}\right| \geq
\left|\tset \setminus \DONE_\EX\right| - (m-1) \geq m$.

It follows that any $p \in \pset_\EX$ that executes action $\act{compNext}_p$ after state $\st_{0}'$, will have its $\nxt{p}$ variable
pointing in a job $x$ with  $\rank{x}{\Pool_\EX} \geq \left\lfloor \left(p-1\right)\cdot 
			 \frac{|\tset \setminus \DONE_\EX|-\left(m-1\right)}{m} \right\rfloor+1$. 
Thus  $\forall p \in \pset_\EX$, $\exists~\st_{p}' \geq \st_{0}'$ in $\EX$ such that $\forall$ states $\st \geq \st_{p}'$,
$\rank{\st.next_p}{\Pool_\EX} \geq \left\lfloor \left(p-1\right)\cdot  \frac{|\tset \setminus \DONE_\EX|-\left(m-1\right)}{m} \right\rfloor+1$. 
Let $\st_{0}''=\max_{p\in \pset_{\EX}}[\st_{p}']$, we have $2$ cases for $p_0$:

{\bf Case A.1)} After $\st_{0}''$, process $p_0$ executes action $\act{compNext}_{p_0}$ and the transition leads to state 
$\st_{1} > \st_{0}''$ such that $\st_{1}.\nxt{p_{0}}=x_0$. Since $\rank{x_{0}}{\Pool_\EX}= 	\left\lfloor \left(p_{0}-1\right)\cdot 
		 \frac{|\tset \setminus \DONE_\EX|-\left(m-1\right)}{m} \right\rfloor+1$ and 
$p_0=\min_{p\in \pset_\EX}[p]$, 
from the previous discussion we have that $\forall
\st \geq \st_{1}$ and $\forall p \in \pset \setminus \left\{p_0\right\}$, $\st.next_p \neq x_0$. Thus when $p_0$ executes 
action $\act{check}_p$ of Fig.\,\ref{fig:IOAtrans} for the first time after state $\st_{1}$, the condition will 
be true, so in some subsequent transition $p_0$ will have to execute action $\act{do}_{p_0,x_0}$, performing 
job $x_0$, which is a contradiction, since after state $\st_{0}$ no jobs are executed.

{\bf Case A.2)} After $\st_{0}''$, process $p_0$ executes action $\act{compNext}_{p_0}$ and the transition leads in state 
$\st_{1} >  \st_{0}''$ such that $\st_{1}.\nxt{p_{0}}>x_0$. 
Since $p_0=\min_{p\in \pset_{\EX}}[p]$, it holds that $\forall x \in \Pool_\EX$ such that $\rank{x}{\Pool_\EX} 
\leq \left\lfloor \left(p_{0}-1\right)\cdot  \frac{|\tset \setminus \DONE_\EX|-\left(m-1\right)}{m} \right\rfloor+1$,
 $\nexists p \in \pset$ such that $\st_{1}.next_p=x$.
Let the transition $\left(\st_{2},\act{compNext}_{p_0},\st_{2}^{'}\right) \in \EX$, where $\st_{2} > \st_{1}$, 
be the first time that action $\act{compNext}_{p_0}$ is executed after state $\st_{1}$. 
We have that $\forall x \in \Pool_\EX$ such that 
$\rank{x}{\Pool_\EX} \leq \left\lfloor \left(p_{0}-1\right)\cdot  \frac{|\tset \setminus \DONE_\EX|-\left(m-1\right)}{m} \right\rfloor+1$, 
$x \notin \st_{2}.\DONE_{p_0}\cup \st_{2}.\TRY_{p_0}$, since from the discussion above we have that
 $\forall \st \geq \st_{1}$ and $\forall p \in \pset_{\EX} \setminus \left\{p_{0}\right\}$, 
$\rank{\st.next_p}{\Pool_\EX} \geq \left\lfloor \left(p-1\right)\cdot \frac{|\tset \setminus \DONE_\EX|-\left(m-1\right)}{m} \right\rfloor+1$. 
Thus $\rank{x_0}{\FREEinS{\st_{2}}{p_0}\setminus \TRYinS{\st_{2}}{p_0}}= \rank{x_0}{\Pool_\EX}=
\left\lfloor \left(p_{0}-1\right)\cdot  \frac{|\tset \setminus \DONE_\EX|-\left(m-1\right)}{m} \right\rfloor+1$. 
As a result, $\st_{2}^{'}.\nxt{p_{0}}=x_0$.
With similar arguments like in case A.1, we can see that job $x_0$ will be performed by process $p_0$, 
which is a contradiction, since after state $\st_{0}$ no jobs are executed.

{\bf Case B $\left| \tset \setminus\DONE_\EX \right| < 2m-1$}: Let $x_{0} \in \Pool_\EX$ be the job such that $\rank{x_{0}}{\Pool_\EX}=p_0$.
Such $x_{0}$ exists since $\beta \geq m$ and $\Pool_\EX \geq \beta$.
It follows that any $p \in \pset_\EX$ that executes action $\act{compNext}_p$ after state $\st_{0}'$, will have its $\nxt{p}$ variable
pointing in a job $x$ with  $\rank{x}{\Pool_\EX} \geq p$. 
Thus  $\forall p \in \pset_\EX$, $\exists~\st_{p}' \geq \st_{0}'$ in $\EX$ such that $\forall$ states $\st \geq \st_{p}'$,
$\rank{\st.next_p}{\Pool_\EX} \geq p$. 
Let $\st_{0}''=\max_{p\in \pset_{\EX}}[\st_{p}']$, we have $2$ cases for $p_0$:

{\bf Case B.1)} After $\st_{0}''$, process $p_0$ executes action $\act{compNext}_{p_0}$ and the transition leads in state 
$\st_{1} > \st_{0}''$ such that $\st_{1}.\nxt{p_{0}}=x_0$. Since $\rank{x_{0}}{\Pool_\EX}= p_0$ and 
$p_0=\min_{p\in \pset_\EX}[p]$, 
from the previous discussion we have that $\forall
\st \geq \st_{1}$ and $\forall p \in \pset \setminus \left\{p_0\right\}$, $\st.next_p \neq x_0$. Thus when $p_0$ executes 
action $\act{check}_p$ of Fig.\,\ref{fig:IOAtrans} for the first time after state $\st_{1}$, the condition will 
be true, so in some subsequent transition $p_0$ will have to execute action $\act{do}_{p_0,x_0}$, performing 
job $x_0$, which is a contradiction, since after state $\st_{0}$ no jobs are executed.

{\bf Case B.2)} After $\st_{0}''$, process $p_0$ executes action $\act{compNext}_{p_0}$ and the transition leads in state 
$\st_{1} >  \st_{0}''$ such that $\st_{1}.\nxt{p_{0}}>x_0$. 
Since $p_0=\min_{p\in \pset_\EX}[p]$, it holds that $\forall x \in \Pool_\EX$ such that $\rank{x}{\Pool_\EX} 
\leq p_0$,
 $\nexists p \in \pset$ such that $\st_{1}.next_p=x$.
Let the transition $\left(\st_{2},\act{compNext}_{p_0},\st_{2}^{'}\right) \in \EX$, where $\st_{2} > \st_{1}$, 
be the first time that action $\act{compNext}_{p_0}$ is executed after state $\st_{1}$. 
We have that $\forall x \in \Pool_\EX$ such that 
$\rank{x}{\Pool_\EX} \leq p_0$, 
$x \notin \st_{2}.\DONE_{p_0}\cup \st_{2}.\TRY_{p_0}$, since from the discussion above we have that
 $\forall \st \geq \st_{1}$ and $\forall p \in \pset_\EX \setminus \left\{p_{0}\right\}$, 
$\rank{\st.next_p}{\Pool_\EX} \geq p$. 
Thus $\rank{x_0}{\FREEinS{\st_{2}}{p_0}\setminus \TRYinS{\st_{2}}{p_0}}= \rank{x_0}{\Pool_\EX}= p_0$. 
As a result, $\st_{2}^{'}.\nxt{p_{0}}=x_0$.
With similar arguments like in case B.1, we can see that job $x_0$ will be performed by process $p_0$, 
which is a contradiction, since after state $\st_{0}$ no jobs are executed.

\end{proof}


We combine the last two lemmas in order to show the main result on the effectiveness of algorithm $\algName$.
\begin{theorem}
\label{thm:effectiveness}
For any $\beta \geq m$, $f \leq m-1$ algorithm $\algName$ has effectiveness $E_{\algName}(n,m,f)=n-\left(\beta + m - 2\right)$.
\end{theorem}
\begin{proof}
From Lemma \ref{lem:noTermination} we have that any finite execution $\EX \in \ex{\algName}$
with $\donum{\EX} \leq n-\left(\beta + m - 1\right)$ can be extended, essentially proving that in 
such executions no process has terminated. Moreover from Lemma \ref{lem:simpleEFF} we have that $\algName$ is
wait free, and thus there exists
no infinite fair execution $\EX \in \ex{\algName}$, such that $\donum{\EX} \leq n-\left(\beta + m - 1\right)$.
Since finite fair executions are executions where all non-failed processes have terminated, from the
above we have that $E_{\algName}(n,m,f)\geq n-\left(\beta + m - 2\right)$.

If all processes but the process with id $m$ fail in an execution $\EX$ in such a way
that $\tset_\EX \cap \Stuck_\EX = \emptyset$ and $|\Stuck_\EX|=m-1$\,(where $\Stuck_\EX$ is defined as in the proof of Lemma~\ref{lem:simpleEFF}), it is easy to see that there exists an adversarial strategy, such that when process $m$ terminates, $\beta+m-2$ jobs have not been performed . 
Such an execution will be a finite fair execution where $n-(\beta+m-2)$ jobs are performed. Thus we have that $E_{\algName}(n,m,f)=n-\left(\beta + m - 2\right)$.
\end{proof}

%
%

\section{Work Complexity Analysis}
\label{sec:work}
In this section we are going to prove that for $\beta \geq 3m^2$ algorithm $\algName$ has work complexity $\mathrm{O}(nm\log n \log m)$.

The main idea of  the proof, is to demonstrate that under the assumption   $\beta \geq 3m^2$,  process {\em collisions} on a job cannot accrue  without making progress in the algorithm.
In order to prove that, we first demonstrate in Lemma\,\ref{lem:collide} that if two different processes $p,q$ set their $\nxt{p}, \nxt{q}$ 
internal variables to the same job $i$ in some $\act{compNext}$ actions, then 
the $\DONE_p$ and $\DONE_q$ sets of the processes, have at least $|q-p|\cdot m$ different elements, given that
$\beta \geq 3m^2$. Next we prove in Lemma\,\ref{lem:collideALL} that if two processes $p,q$ collide three consecutive times, while trying to perform
some jobs, the size of the set $\DONE_p \cup \DONE_q$ that processes $p$ and $q$ know will increase by at
least $|q-p|\cdot m$ elements. This essentially tells us that every three collisions between the same two processes
a significant number of jobs has been performed, and thus enough progress has been made. In order to prove
the above statement, we formally define what we mean by collision in 
Definition\,\ref{def:collision}, and tie such a collision with 
some specific state, the state the collision is detected, so that we have a fixed ``point of reference'' in the execution; and show that
the order collisions are detected in an execution, is consistent with the order
the involved processes attempt to perform the respective jobs in Lemmas\,\ref{lem:collideOrder},\,\ref{lem:collideOrder2}. 
Finally we use Lemma\,\ref{lem:collideALL}, 
in order to prove in Lemma\,\ref{lem:boundCol}, that a process $p$ cannot collide with a process $q$ more than 
$2\left\lceil \frac{n}{m\cdot |q-p|} \right\rceil$ times in any execution.
This is proven by contradiction, showing that if process $p$ collides with process $q$ more than 
$2\left\lceil \frac{n}{m\cdot |q-p|} \right\rceil$ times, there exist states for which the set $|\DONE_p \cup \DONE_q|$ has more than $n$ 
elements which is impossible. Lemma\,\ref{lem:boundCol} is used in order to prove the main result on the work complexity 
of algorithm $\algName$ for $\beta \geq 3m^2$, Theorem\,\ref{thm:work}. We obtain 
Theorem\,\ref{thm:work} by counting the total number of 
collisions that can happen and the cost of each collision. 


We start by defining the notion of \emph{immediate predecessor} transition for a state $\st$ in 
an execution $\EX$. The immediate predecessor is the last transition of a specific action type that precedes state $\st$ in the execution. This is particularly useful in uniquely identifying the  transition with action $\act{compNext}_p$ in an execution, that last 
set a $\nxt{p}$ internal variable to a specific value, given a state $\st$ of interest.

\begin{definition}
\label{def:immediatePredecesor}
We say that transition $\left(\st_{1}, \acts_{1},\st_{1}^{'}\right)$ is an \textbf{\ip{}} of state  $\st_{2}$ in an execution $\EX \in execs(\algName)$ and we write 
$\left(\st_{1}, \acts_{1},\st_{1}^{'}\right) \ipSym \st_{2}$, if  $\st_{1}^{'} < \st_{2}$ and in the execution fragment
$\EX'$  that begins with state $\st_{1}^{'}$ and ends with state $\st_{2}$, 
there exists no action $\acts_{3}= \acts_{1}~$.
\end{definition}

Next we define what a collision between two processes means. We say that process $p$ \emph{collided} with process $q$ in job $i$ at 
state $\st$, if process $p$ attempted to preform job $i$, but was not able to, because it 
detected in state $\st$ that either process $q$ was trying to perform job $i$ or process $q$ has already performed job $i$. 

\begin{definition}
\label{def:collision}
In an execution $\EX \in execs(\algName)$, we say that process $p$ \textbf{collided} with process $q$ in job $i$ at state $\st$,
if (i) there exist in $\EX$ transitions 
$\left(\st_1, \act{compNext}_{p},\st_{1}^{'}\right)$,  
$\left(\stt_1, \act{compNext}_{q},\stt_{1}^{'}\right)$ 
and $\left(\st_2, \act{check}_{p},\st_{2}^{'}\right)$, where 
$\left(\st_1, \act{compNext}_{p},\st_{1}^{'}\right) \ipSym \st_2$, $\stt_1 < \st_2$ and  
$\st_{1}^{'}.\nxt{p} = \stt_{1}^{'}.\nxt{q} = \st_2.\nxt{p} = i$, 
$\st_{1}^{'}.\status{p} = \stt_{1}^{'}.\status{q} = set\_next$,
$\st_{2}^{'}.\status{p} = comp\_next$, 
(ii) in execution fragment $\EX'=\st_1^{'},\ldots,\st_2$ 
either there exists  transition $\left(\st, \act{gatherTry}_{p},\st^{'}\right)$ such that $\st.\q{p}=q, \st.next_q=i$,
or transition $\left(\st, \act{gatherDone}_{p},\st^{'}\right)$ and $j \in \{1,\ldots,n\}$ 
such that $\st.\q{p}=q$, $\st.\pos{p}{q}=j$, $\st.done_{q,j}=i$ and $i \notin \st.\TRY_{p}$.
\end{definition}


\begin{definition}
\label{def:2procCollide}
In an execution $\EX \in execs(\algName)$, we say that processes $p,q$ \textbf{collide} in job $i$ at state $\st$,
if process $p$ collided with process $q$ 
or process $q$ collided with process $p$ in job $i$ at state $\st$, according to 
Definition \ref{def:collision}.
\end{definition}

Next we show that if two processes $p, q$ decide, with some $\act{compNext}$ actions, to perform the same job $i$, then their $\DONE$ sets at the enabling states of those $\act{compNext}$ actions, differ in at-least $|q-p|\cdot m$ elements.

\begin{lemma}
\label{lem:collide}
If  $\beta \geq 3m^2$ and in an execution $\EX \in execs(\algName)$ there exist states $\st_1, \stt_1$ and processes 
$p,q \in \pset$ with $p < q$ such that $\st_1.\nxt{p}=\stt_1.\nxt{q}=i \in \tset$, then there exist transitions 
$\left(\st_2, \act{compNext}_{p},\st_{2}^{'}\right) \ipSym \st_1$, $\left(\stt_2, \act{compNext}_{q},\stt_{2}^{'}\right) \ipSym \stt_1$,  
where $\st_{2}^{'}.\nxt{p}=\stt_{2}^{'}.\nxt{q} = i$, $\st_{2}^{'}.\status{p}=\stt_{2}^{'}.\status{q}=set\_next$ 
and 
$\left|\st_{2}.\DONE_p \cap \overline{\stt_{2}.\DONE_q}\right| > (q-p)\cdot m$ or
$\left|\overline{\st_{2}.\DONE_p} \cap \stt_{2}.\DONE_q\right| > (q-p)\cdot m~$.
\end{lemma}
\begin{proof} 
We will prove this by contradiction. From algorithm $\algName$ there must exist transitions 
$\left(\st_2, \act{compNext}_{p},\st_{2}^{'}\right) \ipSym \st_{1}$ and $\left(\stt_2, \act{compNext}_{q},\stt_{2}^{'}\right) \ipSym \stt_{1}$,
where $\st_{2}^{'}.\nxt{p}=i$ and $\stt_{2}^{'}.\nxt{q}=i$, 
if there exist $\st_1, \stt_1 \in \EX$ and 
$p,q \in \pset$ with $p < q$ such that $\st_1.\nxt{p}=\stt_1.\nxt{q}=i \in \tset$, since those are the transitions
that set $\nxt{p}$ and $\nxt{q}$ to $i$. In order to get a contradiction we assume that 
%
$\left|\st_{2}.\DONE_p \cap \overline{\stt_{2}.\DONE_q}\right|
\leq (q-p)\cdot m$ and $\left|\overline{\st_{2}.\DONE_p} \cap \stt_{2}.\DONE_q\right|
\leq (q-p)\cdot m$.
We will prove that if this is the case, then $\st_{2}^{'}.\nxt{p} \neq \stt_{2}^{'}.\nxt{q}$.

Let $\setA = \tset \setminus \st_{2}.\DONE_p = \FREEinS{\st_{2}}{p}$ and 
$\setB = \tset \setminus \stt_{2}.\DONE_q = \FREEinS{\stt_{2}}{q}$, thus from the contradiction assumption we have that:
$\left|\overline{\setA} \cap \setB \right| \leq (q-p)\cdot m$ and
$\left|\setA \cap \overline{\setB} \right| \leq (q-p)\cdot m$.

It could either be that $|\setA| < |\setB|$ or $|\setA| \geq |\setB|$.

\Par{Case 1 $|\setA| < |\setB|$:}
From the contradiction assumption we have that $\left|\overline{\setA} \cap \setB \right| \leq (q-p)\cdot m$.
Since $\st_2.\FREE_p\setminus\st_2.\TRY_p$ can have up to $m-1$ fewer elements than $\setA$ -- the elements of set $\st_2.\TRY_p$ --
and it can be the case that $\st_2.\TRY_p \cap \stt_2.\TRY_q = \emptyset$, we have:
\begin{equation}
\label{eq:col0}
|\stt_2.\FREE_q \setminus \stt_2.\TRY_q \cap \overline{\st_2.\FREE_p\setminus\st_2.\TRY_p}| \leq m(q-p) + m-1
\end{equation}
Moreover, since $\st_{2}.\FREE_p\setminus\st_2.\TRY_p \subseteq \setA$ and $|\st_{2}.\FREE_p\setminus\st_2.\TRY_p| \geq \beta \geq 3m^2$,
$|\setA| \geq 3m^2$. Similarly $|\setB|  \geq 3m^2$. We have:
\[
(q-1)\frac{|\setB|}{m} = (p-1)\frac{|\setB|}{m} + (q-p)\frac{|\setB|}{m} > (p-1)\frac{|\setA|}{m} + (q-p)\frac{|\setB|}{m} \Rightarrow 
\]
\[
	\Rightarrow (q-1)\frac{|\setB|}{m} > (p-1)\frac{|\setA|}{m} + 3m(q-p) \Rightarrow
\]
\[
\Rightarrow (q-1)\frac{|\setB|}{m} > (p-1)\frac{|\setA|}{m} + (3m-1)(q-p) + (q-p)\Rightarrow 
\]
\[
\Rightarrow (q-1)\frac{|\setB|}{m} > (p-1)\frac{|\setA|}{m} + (3m-1)(q-p) + \frac{(q-p)(m-1)}{m}\Rightarrow 
\]
\begin{equation}
\label{eq:col1}
\left\lfloor (q-1)\frac{|\setB|-(m-1)}{m}\right\rfloor + 1 
\geq \left\lfloor (p-1)\frac{|\setA|-(m-1)}{m}\right\rfloor + 1 + (3m-1)(q-p)
\end{equation}
Since $\st_{2}^{'}.\nxt{p}=\stt_{2}^{'}.\nxt{q}=i$, we have: 
\[
\rank{i}{\st_2.\FREE_p\setminus\st_2.\TRY_p}=\left\lfloor (p-1)\frac{|\setA|-(m-1)}{m}\right\rfloor + 1
\]
\[
\rank{i}{\stt_2.\FREE_q\setminus\stt_2.\TRY_q}=\left\lfloor (q-1)\frac{|\setB|-(m-1)}{m}\right\rfloor + 1
\]
Equation\,\ref{eq:col1} becomes:
\[  
\rank{i}{\stt_2.\FREE_q\setminus\stt_2.\TRY_q} \geq \rank{i}{\st_2.\FREE_p\setminus\st_2.\TRY_p} + (3m-1)(q-p) 
\]
Thus set $\stt_2.\FREE_q\setminus\stt_2.\TRY_q$ must have at least $(3m-1)(q-p)$ more elements with rank less that the rank
of $i$, than set $\st_2.\FREE_p\setminus\st_2.\TRY_p$ does. This is a contradiction since from eq.\,\ref{eq:col0} we have that:
\[
|\stt_2.\FREE_q\setminus\stt_2.\TRY_q \cap \overline{\st_2.\FREE_p\setminus\st_2.\TRY_p}| \leq m(q-p) + m-1
\]

\Par{Case 2 $|\setB| \leq |\setA|$:}
We have that $\left|\overline{\setA} \cap \setB \right| \leq (q-p)\cdot m$ and 
$\left|\setA \cap \overline{\setB} \right| \leq (q-p)\cdot m$ from the contradiction assumption.
Since $\st_2.\FREE_p\setminus\st_2.\TRY_p$ can have up to $m-1$ less elements than $\setA$ -- the elements of set $\st_2.\TRY_p$ --
and it can be the case that $\st_2.\TRY_p \cap \stt_2.\TRY_p = \emptyset$, we have:
\begin{equation}
\label{eq:col2}
|\stt_2.\FREE_q \setminus \stt_2.\TRY_q \cap \overline{\st_2.\FREE_p\setminus\st_2.\TRY_p}| \leq m(q-p) + m-1
\end{equation}
From the contradiction assumption and the case 2 assumption we have that $|\setB| \leq |\setA| \leq |\setB|+(q-p) \cdot m$.
Moreover $|\setA| \geq \beta \geq 3m^2$ and $|\setB| \geq \beta \geq 3m^2$.
We have:
\[
\left(q-1\right)\frac{|\setB|+(q-p)m}{m}=
\left(p-1\right)\frac{|\setB|+(q-p)m}{m}+\left(q-p\right)\frac{|\setB|+(q-p)m}{m}\geq
\]
\[
\geq \left(p-1\right)\frac{|\setA|}{m}+\left(q-p\right)\frac{|\setB|+(q-p)m}{m}\geq
\left(p-1\right)\frac{|\setA|}{m}+3m\left(q-p\right)+\left(q-p\right)^2\Rightarrow
\]
\[
 \Rightarrow \left(q-1\right)\frac{|\setB|}{m} \geq 
 \left(p-1\right)\frac{|\setA|}{m}+3m\left(q-p\right)+\left(q-p\right)^2-(q-1)(q-p)\Rightarrow
\]
\[
\Rightarrow \left(q-1\right)\frac{|\setB|}{m} \geq 
 \left(p-1\right)\frac{|\setA|}{m}+\left(3m-p+1\right)\left(q-p\right)\Rightarrow
\]
\[
\Rightarrow \left(q-1\right)\frac{|\setB|}{m} \geq 
 \left(p-1\right)\frac{|\setA|}{m}+\left(2m+2\right)\left(q-p\right)\Rightarrow
\]
\[
\Rightarrow \left(q-1\right)\frac{|\setB|}{m} \geq 
 \left(p-1\right)\frac{|\setA|}{m}+\left(2m+1\right)\left(q-p\right)+\frac{\left(q-p\right)\left(m-1\right)}{m}\Rightarrow
\]
\begin{equation}
\label{eq:col3}	 
\left\lfloor (q-1)\frac{|\setB|-(m-1)}{m}\right\rfloor + 1 
\geq \left\lfloor (p-1)\frac{|\setA|-(m-1)}{m}\right\rfloor + 1 + (2m+1)(q-p)
\end{equation}
Since $\st_{2}^{'}.\nxt{p}=\stt_{2}^{'}.\nxt{q}=i$, we have:
\[
\rank{i}{\st_2.\FREE_p\setminus\st_2.\TRY_p}=\left\lfloor (p-1)\frac{|\setA|-(m-1)}{m}\right\rfloor + 1
\]
\[
\rank{i}{\stt_2.\FREE_q\setminus\stt_2.\TRY_q}=\left\lfloor (q-1)\frac{|\setB|-(m-1)}{m}\right\rfloor + 1
\]
Equation\,\ref{eq:col3} becomes:
\[
\rank{i}{\stt_2.\FREE_q\setminus\stt_2.\TRY_q} \geq 
\rank{i}{\st_2.\FREE_p\setminus\st_2.\TRY_p} + (2m+1)(q-p)
\] 
Thus set $\stt_2.\FREE_q\setminus\stt_2.\TRY_q$ must have at least $(2m+1)(q-p)$ more elements with rank less that the rank
of $i$, than set $\st_2.\FREE_p\setminus\st_2.\TRY_p$. This is a contradiction since from 
eq.\,\ref{eq:col2} we have that:
\[
|\stt_2.\FREE_q\setminus\stt_2.\TRY_q \cap \overline{\st_2.\FREE_p\setminus\st_2.\TRY_p}| \leq m(q-p) + m-1
\]
\end{proof}

Next we show that if a process $p$ detects consecutive collisions with process $q$, 
the processes $p, q$ attempted to perform the jobs associated with the collisions 
in the same order and the
order process $p$ detects the collisions according to Definition\,\ref{def:collision} is the same as the order processes 
$p, q$ attempted to perform the jobs.  

In the proofs that follow, for a state $\st$ in execution $\EX$ we define as $\st.\DONE$ the following set: 
$\st.\DONE = \left\{i \in \tset | \exists p \in \pset~\text{and}~j\in\{1,\ldots,n\}:\st.done_p(j)=i\right\}$.

\begin{lemma}
\label{lem:collideOrder}
In an execution $\EX \in execs(\algName)$ for any $\beta \geq m$ if there exist processes $p,q$, jobs $i_1,i_2 \in \tset$
and states $\tilde{\st}_1 < \tilde{\st}_2$
such that process $p$ collided with process $q$ in job $i_1$ at state $\tilde{\st}_1$ and in job $i_2$  at state $\tilde{\st}_2$
 according to Definition \ref{def:collision},
then there exist transitions  $\left(\st_1, \act{compNext}_{p},\st_{1}^{'}\right) \ipSym \tilde{\st}_1$, $\left(\st_2, \act{compNext}_{p},\st_{2}^{'}\right) \ipSym \tilde{\st}_2$ and
$\left(\stt_1, \act{compNext}_{q},\stt_{1}^{'}\right)$, $\left(\stt_2, \act{compNext}_{q},\stt_{2}^{'}\right)$ where
$\st_{1}^{'}.\nxt{p}=\stt_{1}^{'}.\nxt{q}=i_1$, 
$\st_{2}^{'}.\nxt{p}=\stt_{2}^{'}.\nxt{q}=i_2$,
$\st_{1}^{'}.\status{p}=\st_{2}^{'}.\status{p}=\stt_{1}^{'}.\status{q}=\stt_{2}^{'}.\status{q}=set\_next$ 
such that: 
\begin{center}
$\st_1 < \st_2$ and $\stt_1  < \stt_2~$.
\end{center}
\end{lemma}
\begin{proof} 
From Definition\,\ref{def:collision} we have that there exist transitions 
$\left(\st_1, \act{compNext}_{p},\st_{1}^{'}\right)$, $\left(\st_2, \act{compNext}_{p},\st_{2}^{'}\right)$ with 
$\st_{1}^{'}.\nxt{p}=i_1$, $\st_{2}^{'}.\nxt{p}=i_2$, $\st_{1}^{'}.\status{p}=\st_{2}^{'}.\status{p}=set\_next$,
and there exists no action $\acts_{1} = \act{compNext}_p$ for which $\st_1 < \acts_{1} < \tilde{\st}_1$ or
$\st_2 < \acts_{1} < \tilde{\st}_2$. From the latter and the fact that  $\tilde{\st}_1 < \tilde{\st}_2$, 
it must be the case that  $\st_1 < \tilde{\st}_1 < \st_2  < \tilde{\st}_2$. 
Furthermore from Definition  \ref{def:collision} we have that there exist transitions 
$\left(\stt_1, \act{compNext}_{q},\stt_{1}^{'}\right)$, $\left(\stt_2, \act{compNext}_{q},\stt_{2}^{'}\right)$ with
$\stt_{1}^{'}.\nxt{q}=i_1$, $\stt_{2}^{'}.\nxt{q}=i_2$, $\stt_{1}^{'}.\status{q}=\stt_{2}^{'}.\status{q}=set\_next$,
such that $\stt_{1}^{'} < \tilde{\st}_1$ and $\stt_{2}^{'} < \tilde{\st}_2$. We can pick those transitions in
$\EX$ in such a way that there exists no other transition between $\stt_{1}^{'}$ and $\tilde{\st}_1$ that sets 
$\nxt{q}$ to $i_1$ and similarly there exists no other transition 
between $\stt_{2}^{'}$ and $\tilde{\st}_2$ that sets $\nxt{q}$ to $i_2$.
We need to prove now that   $\stt_1  < \stt_2$. We will prove this by contradiction. 

Let $\stt_2 < \stt_1$. Since
$\stt_{1}^{'} < \tilde{\st}_1$, we have that $\stt_2 < \stt_1 < \stt_{1}^{'} < \tilde{\st}_1 < \st_2 < \tilde{\st}_2$.
Since from Definition\,\ref{def:collision} either $\tilde{\st}_1.next_q = i_1$ or there exists $j \in \{1,\ldots,n\}$ 
such that $\tilde{\st}_1.done_{q,j} = i_1$, it must be the case that $\tilde{\st}_2.\status{p} = gather\_done$,
$\tilde{\st}_2.\q{p} = q$ and there exists $j' \in \{1,\ldots,n\}$ such that $\tilde{\st}_2.done_{q,j'} = i_2$. 
Essentially, it must be that case that process $q$ performed job $i_2$ after transition $\left(\stt_2, \act{compNext}_{q},\stt_{2}^{'}\right)$.
This means that there exists transition $\left(\stt_3, \act{done}_{q},\stt_{3}^{'}\right)$ and $j' \in \{1,\ldots,n\}$
such that $\stt_{3}^{'}.done_{q,j'} = i_2$ and  
$\stt_2 < \stt_{3}^{'} <\stt_1 < \stt_{1}^{'} < \tilde{\st}_1 < \st_2 < \tilde{\st}_2$.

If $\tilde{\st}_1.\status{p}=gather\_try$ then from algorithm $\algName$ we have that  
$\tilde{\st}_1.\DONE \subseteq \st_2.\DONE_p$, since actions $\act{gatherTry}_p$ 
are followed by actions $\act{gatherDone}_p$ before any action $\act{setNext}_p$ takes place. 
As a result $i_2 \in \st_2.\DONE_p$, which is a contradiction since
$\left(\st_2, \act{compNext}_{p},\st_{2}^{'}\right) \notin trans(\algName)$ if $i_2 \in \st_2.\DONE_p$ and 
$\st_{2}^{'}.\nxt{p} = i_2$, $\st_{2}^{'}.\status{p} = set\_next$.

If $\tilde{\st}_1.\status{p}=gather\_done$ then from algorithm $\algName$ we have that  
$\tilde{\st}_1.\q{p}= q$ and there exists $j \in \{1,\ldots,n\}$ such that
$\tilde{\st}_1.\pos{p}{q}= j$ and $\tilde{\st}_1.done_{q,j}= i_1$. Since
$\stt_2 < \stt_{3}^{'} <\stt_1 < \stt_{1}^{'} < \tilde{\st}_1 < \st_2 < \tilde{\st}_2$
it must be the case that $j'< j$ and as a result $i_2 \in \tilde{\st}_1.\DONE_{p}$.
Clearly  $\tilde{\st}_1.\DONE_{p} \subseteq \st_2.\DONE_p$, which is a contradiction since
$\left(\st_2, \act{compNext}_{p},\st_{2}^{'}\right) \notin trans(\algName)$ if $i_2 \in \st_2.\DONE_p$ and 
$\st_{2}^{'}.\nxt{p} = i_2$, $\st_{2}^{'}.\status{p} = set\_next$. 
\end{proof}

Next we show that if two consecutive collisions take place between processes $p,q$, and $p$ detects the one collision and $q$ the other, the processes
$p, q$ attempted to perform the jobs associated with the collisions in the same order and
the order in which  the processes detect the collisions according to 
Definition\,\ref{def:collision} is the same as the order the processes 
$p, q$ attempted to perform the jobs.

\begin{lemma}
\label{lem:collideOrder2}
In an execution $\EX \in execs(\algName)$ for any $\beta \geq m$ if there exist processes $p,q$, jobs $i_1,i_2 \in \tset$
and states $\tilde{\st}_1 < \tilde{\st}_2$
such that process $p$ collided with process $q$ in job $i_1$ at state $\tilde{\st}_1$ and 
process $q$ collided with process $p$ in job $i_2$  at state $\tilde{\st}_2$
 according to Definition \ref{def:collision},
then there exist transitions  
$\left(\st_1, \act{compNext}_{p},\sk{\st_{1}^{'}}\right) \ipSym \tilde{\st}_1$, 
$\left(\stt_2, \act{compNext}_{q},\stt_{2}^{'}\right) \ipSym \tilde{\st}_2$  and 
$\left(\st_2, \act{compNext}_{p},\st_{2}^{'}\right)$,
$\left(\stt_1, \act{compNext}_{q},\stt_{1}^{'}\right)$,  where
$\st_{1}^{'}.\nxt{p}=\stt_{1}^{'}.\nxt{q}=i_1$, 
$\st_{2}^{'}.\nxt{p}=\stt_{2}^{'}.\nxt{q}=i_2$,
$\st_{1}^{'}.\status{p}=\st_{2}^{'}.\status{p}=\stt_{1}^{'}.\status{q}=\stt_{2}^{'}.\status{q}=set\_next$ 
such that: 
\begin{center}
$\st_1 < \st_2$ and $\stt_1  < \stt_2~$.
\end{center}
\end{lemma}
\begin{proof} 
From Definition  \ref{def:collision} we have that there exist transitions 
$\left(\st_1, \act{compNext}_{p},\st_{1}^{'}\right)$, $\left(\st_2, \act{compNext}_{p},\st_{2}^{'}\right)$ with 
$\st_{1}^{'}.\nxt{p}=i_1$, $\st_{2}^{'}.\nxt{p}=i_2$, $\st_{1}^{'}.\status{p}=\st_{2}^{'}.\status{p}=set\_next$,
and there exists no action $\acts_{1} = \act{compNext}_p$ for which $\st_1 < \acts_{1} < \tilde{\st}_1$. 
Furthermore from Definition  \ref{def:collision} we have that there exist transitions 
$\left(\stt_1, \act{compNext}_{q},\stt_{1}^{'}\right)$, $\left(\stt_2, \act{compNext}_{q},\stt_{2}^{'}\right)$ with
$\stt_{1}^{'}.\nxt{q}=i_1$, $\stt_{2}^{'}.\nxt{q}=i_2$, $\stt_{1}^{'}.\status{q}=\stt_{2}^{'}.\status{q}=set\_next$,
and there exists no action $\acts_{2} = \act{compNext}_q$ for which $\stt_2 < \acts_{2} < \tilde{\st}_2$.
From the later and the fact that  $\tilde{\st}_1 < \tilde{\st}_2$, 
it must be the case that  
$\stt_1 < \stt_2  < \tilde{\st}_2$.
We can pick the transitions that are enabled by states $\stt_1$ and $\st_2$ in 
$\EX$ in such a way that there exists no other transition between $\stt_{1}^{'}$ and $\tilde{\st}_1$ that sets 
$\nxt{q}$ to $i_1$ and similarly there exists no other transition 
between $\st_{2}^{'}$ and $\tilde{\st}_2$ that sets $\nxt{p}$ to $i_2$.
We need to prove now that   $\st_1  < \st_2$. We will prove this by contradiction. 

Let $\st_2 < \st_1$. From algorithm $\algName$ and Definition  \ref{def:collision} there exist 
transitions $\left(\st_3, \act{setNext}_{p},\st_{3}^{'}\right)$, 
and
$\left(\stt_3, \act{setNext}_{q},\stt_{3}^{'}\right)$, where $\st_{3}^{'}.next_p = i_2$, 
$\stt_{3}^{'}.next_q = i_1$ and  $\st_2 < \st_{3}^{'} < \st_1$, $\stt_1 < \stt_{3}^{'} < \stt_2$.
There are 2 cases, either  $\st_{3}^{'} < \stt_{3}^{'}$ or $\stt_{3}^{'} < \st_{3}^{'}$.

\Par{Case 1 $\st_{3}^{'} < \stt_{3}^{'}$:}
We have that $\st_{3}^{'} < \stt_{3}^{'} < \stt_2$ and $\left(\stt_2, \act{compNext}_{q},\stt_{2}^{'}\right)$, where 
$\stt_{2}^{'}.\nxt{q}=i_2$ and $\stt_{2}^{'}.\status{q}=set\_next$ which means that $i_2 \notin 
\stt_{2}.\TRY_{q} \cup \stt_{2}.\DONE_{q}$. This is a contradiction since the $\stt_{2}.\TRY_{q}$ and $\stt_{2}.\DONE_{q}$
are computed by actions $\act{gatherTry}_q$ and $\act{gatherDone}_q$ that are preceded by state $\st_{3}^{'}$.
Either $i_2 \in \stt_{2}.\TRY_{q}$ or a new action $\act{setNext}_p$  
took place before the $\act{gatherTry}_q$ actions. In the latter case, if there is 
a transition $\left(\st_4, \act{done}_{p},\st_{4}^{'}\right)$, where 
$\st_{4}.next_p = i_2$,  before the action $\act{setNext}_p$, it must be the case that 
$i_2 \in \stt_{2}.\DONE_{q}$.
If there exists no such transition we have again a contradiction since we cannot
have a collision in job $i_2$ at state $\tilde{\st}_2$ as defined in Definition\,\ref{def:collision}.

\Par{Case 2 $\stt_{3}^{'} < \st_{3}^{'}$:}
We have that $\stt_{3}^{'} < \st_{3}^{'} < \st_1$ and $\left(\st_1, \act{compNext}_{p},\st_{1}^{'}\right)$, where 
$\st_{1}^{'}.\nxt{p}=i_1$ and $\st_{1}^{'}.\status{p}=set\_next$ which means that $i_1 \notin 
\st_{1}.\TRY_{p} \cup \st_{1}.\DONE_{p}$. This is a contradiction since the $\st_{1}.\TRY_{p}$ and $\st_{1}.\DONE_{p}$
sets are computed by $\act{gatherTry}_p$ and $\act{gatherDone}_p$ actions that are preceded by state $\stt_{3}^{'}$.
Either $i_1 \in \st_{1}.\TRY_{p}$ or a new action $\act{setNext}_q$  
took place before the $\act{gatherTry}_p$ actions. In the latter case, if there is 
a transition $\left(\stt_4, \act{done}_{q},\stt_{4}^{'}\right)$, where 
$\stt_{4}.next_q = i_1$,  before the action $\act{setNext}_q$, it must be the case that 
$i_1 \in \st_{1}.\DONE_{p}$.
If there exists no such transition we have again a contradiction since we cannot
have a collision in job $i_1$ at state $\tilde{\st}_1$ as defined in Definition\,\ref{def:collision}.
%
\end{proof}

Next we show that if 2 processes $p,q \in \pset$ 
\emph{collide} \sk{three} times, their 
$\DONE$ sets at the \sk{third} collision will contain at least $m\cdot (q-p)$ more jobs than they did at the first collision.
This will allow us to find an upper bound on the collisions a process may participate in. It is possible that
both processes become aware of a collision or only one of them does while the other one successfully completes the job.

\begin{lemma}
\label{lem:collideALL}
If $\beta \geq 3m^2$ and in an execution $\EX \in execs(\algName)$ there exist processes $p\neq q$, jobs $i_1,i_2,i_3 \in \tset$
and states $\tilde{\st}_1 < \tilde{\st}_2 < \tilde{\st}_3$
such that process $p,q$ collide in job $i_1$ at state $\tilde{\st}_1$, in job $i_2$  at state $\tilde{\st}_2$
 and in job $i_3$  at state $\tilde{\st}_3$ according to Definition \ref{def:2procCollide},
then there exist states $\st_1 < \st_3$ and $\stt_1  < \stt_3$ such that:
\[
 \st_{1}.\DONE_p \cup \stt_{1}.\DONE_q \subseteq \st_{3}.\DONE_p \cap \stt_{3}.\DONE_q  
\]
\[
\left|\st_{3}.\DONE_p \cup \stt_{3}.\DONE_q\right| - \left|\st_{1}.\DONE_p \cup \stt_{1}.\DONE_q \right| \geq m\cdot |q-p|
\]
\end{lemma}
\begin{proof} 
From Definitions \ref{def:collision}, \ref{def:2procCollide} we have that there exist transitions
$\left(\st_1, \act{compNext}_{p},\st_{1}^{'}\right)$, $\left(\st_2, \act{compNext}_{p},\st_{2}^{'}\right)$,
$\left(\st_3, \act{compNext}_{p},\st_{3}^{'}\right)$ and 
$\left(\stt_1, \act{compNext}_{q},\stt_{1}^{'}\right)$, $\left(\stt_2, \act{compNext}_{q},\stt_{2}^{'}\right)$,
$\left(\stt_3, \act{compNext}_{q},\stt_{3}^{'}\right)$,
where $\st_{1}^{'}.\nxt{p}=\stt_{1}^{'}.\nxt{q}=i_1$, 
$\st_{2}^{'}.\nxt{p}=\stt_{2}^{'}.\nxt{q}=i_2$, $\st_{3}^{'}.\nxt{p}=\stt_{3}^{'}.\nxt{q}=i_3$,
$\st_{1}^{'}.\status{p}=\st_{2}^{'}.\status{p}=\st_{3}^{'}.\status{p}=
\stt_{1}^{'}.\status{q}=\stt_{2}^{'}.\status{q}=\stt_{3}^{'}.\status{q}=set\_next$ and
$\st_1 < \tilde{\st}_1$, $\stt_1 < \tilde{\st}_1$, $\st_2 < \tilde{\st}_2$, $\stt_2 < \tilde{\st}_2$, and 
$\st_3 < \tilde{\st}_3$, $\stt_3 < \tilde{\st}_3$. We pick from $\EX$ the transitions 
$\left(\st_1, \act{compNext}_{p},\st_{1}^{'}\right)$, $\left(\stt_1, \act{compNext}_{q},\stt_{1}^{'}\right)$,
in such a way that there exists no other $\act{compNext}_{p}$ , $\act{compNext}_{q}$ between states 
$\st_1$, $\tilde{\st}_1$ respectively $\stt_1$, $\tilde{\st}_1$ that sets $\nxt{p}$ respectively 
$\nxt{q}$ to $i_1$. We can pick in a similar manner the transitions for jobs $i_2$, $i_3$. 
From Lemmas
\ref{lem:collideOrder}, \ref{lem:collideOrder2} and Definitions \ref{def:collision}, \ref{def:2procCollide}
we have that $\st_1 < \st_2 < \st_3$ and $\stt_1 < \stt_2 < \stt_3$. 
We will first prove that:
\[
 \st_{1}.\DONE_p \cup \stt_{1}.\DONE_q \subseteq \st_{3}.\DONE_p \cap \stt_{3}.\DONE_q  
\]
From algorithm $\algName$ we have that there exists in $\EX$ transitions 
$\left(\st_4, \act{setNext}_{p},\st_{4}^{'}\right)$, 
$\left(\stt_4, \act{setNext}_{q},\stt_{4}^{'}\right)$ with 
$\st_{4}^{'}.next_p = i_2$, $\stt_{4}^{'}.next_q = i_2$ and there exist no action 
$\acts_1 = compNext_p$, such that $\st_2^{'} < \acts_1 <\st_{4}$, and 
no action $\acts_2 = compNext_q$ , such that $\stt_2^{'} < \acts_2 <\stt_{4}$.
We need to prove that $\stt_1 < \st_4$ and $\st_1 < \stt_4$.

We start by proving that $\stt_1 < \st_4$. 
In order to get a contradiction we 
assume that $\st_4 < \stt_1$. From algorithm $\algName$ we have that there exists in $\EX$ transition
$\left(\stt_5, \act{gatherTry}_{q},\stt_{5}^{'}\right)$, with $\stt_5.\q{q}=p$, and
there exists no action $\acts_2 = compNext_q$, such that $\stt_5^{'} < \acts_2 <\stt_{2}$.
We have that $\st_4 < \stt_1 < \stt_5^{'} <\stt_{2}$ and $i_2 \notin \stt_{2}.\TRY_q \cup \stt_{2}.\DONE_q$.
If $\stt_5.next_p = i_2$ we  have a contradiction since $i_2 \in \st_{2}.\TRY_{q}$.
If $\stt_5.next_q \neq i_2$ there exists an action $\acts_{3} = \act{setNext}_p$ in $\EX$, such that 
$\st_{4} < \acts_{3} < \stt_5$. If this $\acts_{3} = \act{setNext}_p$ is preceded by transition 
$\left(\st_5, \act{done}_{p},\st_{5}^{'}\right)$ with $\st_5.\nxt{p} = i_2$, we have a contradiction since 
$i_2 \in \stt_{5}.\DONE$ and $\stt_{2}.\DONE_{q}$ is computed by $\act{gatherDone}_q$ actions that are preceded
by state $\stt_{5}$, which results in $i_2 \in \stt_{2}.\DONE_{q}$.
If there exists no such transition we have again a contradiction since we cannot
have a collision in job $i_2$ at state $\tilde{\st}_2$ as defined in Definition\,\ref{def:collision}.

The case $\st_1 < \stt_4$ is symmetric and can be proved with similar arguments.

From the discussion above we have that $\stt_1 < \st_4$, thus $\stt_1.\DONE_q \subseteq \st_4.\DONE$. Moreover 
$\st_{3}.\DONE_{p}$ is computed by $\act{gatherDone}_p$ actions that are preceded
by state $\st_{4}$, from which we have that $\stt_1.\DONE_q \subseteq \st_{3}.\DONE_{p}$. 
Since $\st_1 < \st_3$ it holds that $\st_1.\DONE_p \subseteq \st_{3}.\DONE_{p}$, thus we have that
$\st_1.\DONE_p \cup \stt_1.\DONE_q \subseteq \st_{3}.\DONE_{p}$.
From $\st_1 < \stt_4$, with similar arguments as before, we can prove that $\st_1.\DONE_p \cup \stt_1.\DONE_q \subseteq \stt_{3}.\DONE_{q}$,
which gives us that: 
\[\st_1.\DONE_p \cup \stt_1.\DONE_q \subseteq \st_{3}.\DONE_{p} \cap \stt_{3}.\DONE_{q}\]

Now it only remains to prove that:
\[
\left|\st_{3}.\DONE_p \cup \stt_{3}.\DONE_q\right| - \left|\st_{1}.\DONE_p \cup \stt_{1}.\DONE_q \right| > m\cdot |q-p|
\]
If $p<q$ from Lemma \ref{lem:collide} we have that 
$\left|\st_{3}.\DONE_p \cap \overline{\stt_{3}.\DONE_q}\right| > (q-p)m$ or 
$\left|\overline{\st_{3}.\DONE_p} \cap \stt_{3}.\DONE_q\right| > (q-p)m$ . Since 
 $\st_1.\DONE_p \cup \stt_1.\DONE_q \subseteq \st_{3}.\DONE_{p} \cap \stt_{3}.\DONE_{q}$, we have that: 
\[
\left|\st_{3}.\DONE_p \cup \stt_{3}.\DONE_q\right| - \left|\st_{1}.\DONE_p \cup \stt_{1}.\DONE_q \right| > (q-p)\cdot m
\]
If $q<p$ with similar arguments we have that:
\[
\left|\st_{3}.\DONE_p \cup \stt_{3}.\DONE_q\right| - \left|\st_{1}.\DONE_p \cup \stt_{1}.\DONE_q \right| > (p-q)\cdot m
\]
Combining the above we have:
\[
\left|\st_{3}.\DONE_p \cup \stt_{3}.\DONE_q\right| - \left|\st_{1}.\DONE_p \cup \stt_{1}.\DONE_q \right| > m\cdot |q-p|
\]
\end{proof}

Next we prove that a process $p$ cannot collide with a process $q$ more than 
$2\left\lceil \frac{n}{m\cdot |q-p|} \right\rceil$ times in any execution.

\begin{lemma}
\label{lem:boundCol}
If  $\beta \geq 3m^2$  there exists no execution $\EX \in execs(\algName)$ at which process $p$ collided with process $q$ in more
than $2\left\lceil \frac{n}{m|q-p|}\right\rceil$ states according to Definition \ref{def:collision}.
\end{lemma}

\begin{proof} 
Let execution $\EX \in execs(\algName)$ be an execution  at which process $p$ collided with process $q$ in at least
$2\left\lceil \frac{n}{m|q-p|}\right\rceil+1$ states. Let us examine the first  
$2\left\lceil \frac{n}{m|q-p|}\right\rceil+1$ such states. Let those states be
$\tilde{\st}_{1} < \tilde{\st}_{2} < \ldots < \tilde{\st}_{2\left\lceil \frac{n}{m|q-p|}\right\rceil} 
< \tilde{\st}_{2\left\lceil \frac{n}{m|q-p|}\right\rceil+1}$.
From Lemma \ref{lem:collideOrder} we have that there exists states  
$\st_{1} < \st_{2} < \ldots < \st_{2\left\lceil \frac{n}{m|q-p|}\right\rceil} 
< \st_{2\left\lceil \frac{n}{m|q-p|}\right\rceil+1}$ that enable the $compNext_p$ actions and
states $\stt_{1} < \stt_{2} < \ldots < \stt_{2\left\lceil \frac{n}{m|q-p|}\right\rceil} 
< \stt_{2\left\lceil \frac{n}{m|q-p|}\right\rceil+1}$ that enable the $compNext_q$ actions
that lead to the collisions in states 
$\tilde{\st}_{1} < \tilde{\st}_{2} < \ldots < \tilde{\st}_{2\left\lceil \frac{n}{m|q-p|}\right\rceil} 
< \tilde{\st}_{2\left\lceil \frac{n}{m|q-p|}\right\rceil+1}$. Then from  Lemma \ref{lem:collideALL}
we have that $\forall i \in \left\{1,\ldots ,\left\lceil \frac{n}{m|q-p|}\right\rceil\right\}$:
\[
\left|\st_{2i+1}.\DONE_p \cup \stt_{2i+1}.\DONE_q\right| - \left|\st_{2i-1}.\DONE_p \cup \stt_{2i-1}.\DONE_q \right| > m|q-p|
\]
\[
\left|\st_{2i+1}.\DONE_p \cup \stt_{2i+1}.\DONE_q\right| - \left|\st_{1}.\DONE_p \cup \stt_{1}.\DONE_q \right| > im|q-p|
\]
\begin{equation}
\label{eq:imCollisions}
\left|\st_{2i+1}.\DONE_p \cup \stt_{2i+1}.\DONE_q\right| > im|q-p|
\end{equation}
From eq.\,\ref{eq:imCollisions} we have that:
\begin{equation}
\label{eq:contraCollision}
\left|\st_{2\left\lceil \frac{n}{m|q-p|}\right\rceil+1}.\DONE_p \cup \stt_{2\left\lceil \frac{n}{m|q-p|}\right\rceil+1}.\DONE_q\right|
> m|q-p|\left\lceil \frac{n}{m|q-p|}\right\rceil \geq n
\end{equation}
Equation \ref{eq:contraCollision} leads to a contradiction since $\st_{2\left\lceil \frac{n}{m|q-p|}\right\rceil+1}.\DONE_p \cup \stt_{2\left\lceil \frac{n}{m|q-p|}\right\rceil+1}.\DONE_q \subseteq \tset$ and $|\tset| = n$.

\end{proof}

Finally we are ready to prove the main theorem on the work complexity of algorithm $\algName$
for $\beta \geq 3m^2$ .

\begin{theorem}
\label{thm:work}
If $\beta \geq 3m^2$ algorithm $\algName$ has work complexity $W_{\algName} = \BigO{nm\log n \log m}$.
\end{theorem}
\begin{proof} 
We start with the observation that in any execution $\EX$ of algorithm $\algName$, if there exists process
$p$, job $i$, transition $\left(\st_1, \act{done}_{p},\st_1^{'}\right)$ and $j \in \{1,\ldots,n\}$ such that 
$\st_1.\pos{p}{p}=j$, $\st_1.\nxt{p}=i$, for any process $q \neq p$ there exists at most one transition 
$\left(\stt_1, \act{gatherDone}_{q},\stt_1^{'}\right)$ in $\EX$, with $\stt_1.\q{q}=p$, $\stt_1.\pos{q}{p}=j$
and  $\stt_1 \geq \st_1$. Such transition performs exactly one read operation
from the shared memory, one insertion at the set $\DONE_q$ and one removal from the set $\FREE_q$, thus such a transition 
costs $\BigO{\log n}$ work. Clearly there exist 
at most $m-1$ such transitions for each $\act{done}_{p}$.
From Lemma \ref{lem:simpleCorrect} for all processes there can be at most $n$ actions 
$\act{done}_{p}$ in any execution $\EX$ of algorithm $\algName$. Each  $\act{done}_{p}$
action performs one write operation in shared memory, one insertion at the set $\DONE_p$ and one removal from the set $\FREE_p$,
thus such an action has cost $\BigO{\log n}$ work.
Furthermore any $\act{done}_{p}$ is preceded by $m-1$
$\act{gatherTry}_p$ read actions that read the $next$ array and each add at most one element to the 
set $\TRY_p$ with cost $\BigO{\log n}$ and $m-1$ $\act{gatherDone}_p$ read 
actions that do not add elements in the $\DONE_p$ set.
Note that we have already counted the $\act{gatherDone}_{p}$
read actions that result in adding jobs at the $\DONE_p$ set. 
Finally any $\act{done}_{p}$ action is preceded by one $\act{compNext}_{p}$ action. This
action is dominated by the cost of the $rank(\FREE_p, \TRY_p, i)$ function. 
If the sets $\FREE_p$, $\TRY_p$ are represented with some efficient tree structure 
like \emph{red-black tree} or some variant of \emph{B-tree}\,\cite{Bayer72,GuibasS78} 
that allows insertion, deletion and search of an element in $\BigO{\log n}$, 
an invocation of function $rank(\FREE_p, \TRY_p, i)$  costs $\BigO{m \log n}$ work. 
That gives us a total of $\BigO{n m \log n}$ work associated with the $\act{done}_{p}$ actions.

If a process $p$ collided with a process $q$ in job $i$ at state $\st$, we have an
extra $\act{compNext}_p$ action, 
$m-1$ extra $\act{gatherTry}_p$ read actions and insertions in the $\TRY_p$ set and $m-1$ $\act{gatherDone}_p$ read 
actions that do not add elements in the $\DONE_p$ set. Thus each collision costs $\BigO{m \log n}$ work. 
Since $\beta \geq 3m^2$ from Lemma $\ref{lem:boundCol}$ for two distinct processes $p, q$ 
we have that in any execution $\EX$ of algorithm $\algName$ there exist less than 
$2\left\lceil \frac{n}{m|q-p|}\right\rceil$ collisions. 
For process $p$ if we count all such collisions with any other process $q$ we get:
\[
\sum_{q \in \pset-\{p\}}2\left\lceil \frac{n}{m|q-p|}\right\rceil \leq 
2(m-1) + \frac{2n}{m}\sum_{q \in \pset-\{p\}} \frac{1}{|q-p|} \leq 
\]
\begin{equation} \leq 
2(m-1) + \frac{4n}{m}\sum_{i =1}^{\left\lceil \frac{m}{2}\right\rceil} \frac{1}{i} \leq
2(m-1) + \frac{4n}{m} \log m 
\end{equation}
If we count the total number of collisions for all the $m$ processes we get that if 
$\beta \geq 3m^2$ in any execution of algorithm $\algName$ there can be at most $2m^2 + 4n \log m < 4(n+1) \log m$ 
collisions (since $n > \beta$).
Thus collisions cost $\BigO{n m \log n \log m}$ work.
Finally any process $p$ that fails may add in the work complexity less than $\BigO{m \log n}$ work from its $\act{compNext}_p$ action
and from reads (if the process
fails without performing a $\act{done}_p$ action after its latest $\act{compNext}_p$ action).
So for the work complexity of algorithm $\algName$ if  $\beta \geq 3m^2$ we have that 
$W_{\algName} = \mathrm{O}(n m \log n \log m)$.
\end{proof}

\section{An Asymptotically Work Optimal Algorithm}
\label{sec:ItAlg}
We demonstrate how to  solve the at-most-once problem 
with effectiveness $n-\BigO{m^2 \log n \log m}$ and work complexity $\BigO{n +  m^{(3+\epsilon)} \log n}$,
for any constant $\epsilon > 0$, such that $1 / \epsilon$ is a positive integer,
when $m = \BigO{\sqrt[3]{n}}$, using algorithm  $\algName$ with $\beta = 3m^2$.
Algorithm $\ItAlg{\epsilon}$, presented in Fig.\,\ref{fig:IterKK}, performs iterative calls to a variation of algorithm
$\algName$, called $\ItStepName$. $\ItAlg{\epsilon}$ has $3 + 1 / \epsilon$ distinct 
matrices $done$ and vectors $next$
in shared memory, with different granularities. One $done$ matrix, 
stores the regular jobs performed, while the remaining $2 + 1 / \epsilon$
matrices store \emph{super-jobs}. Super-jobs are groups of consecutive jobs. 
From them, one stores super-jobs of size $m \log n \log m$, while the remaining $1 + 1 / \epsilon$ matrices, 
store super-jobs of size  $m^{1-i\epsilon} \log n \log^{1+i} m$ for $i \in \{1, \ldots, 1/\epsilon \}$. The $3 + 1 / \epsilon$ distinct vectors $next$ are used in a similar
way as the matrices $done$.
\begin{figure}[!ht]
	\hrule
	\FF
	{\scriptsize \noindent{\bf $\ItAlg{\epsilon}$ for process $p$:}} \\
	{\scriptsize
	  00 ~ $\size{1} \gets 1 $ \\
	  01 ~ $\size{2} \gets m\log n \log m \\$
	  02 ~ $\FREE_p \gets \mapf{\tset, \size{1} , \size{2}}$ \\
	  03 ~ $\FREE_p \gets \ItStep{\FREE_p, \size{2}}$\\
	  04 ~ {\bf for}$(i \gets 1, i \leq 1/\epsilon, i++)$\\
	  05 ~ \T $\size{1} \gets \size{2} $ \\
	  06 ~ \T $\size{2} \gets m^{1-i\epsilon} \log n \log^{1+i} m  \\$
	  07 ~ \T $\FREE_p \gets \mapf{\FREE_p, \size{1}, \size{2} }$ \\
	  08 ~ \T $\FREE_p \gets \ItStep{\FREE_p, \size{2}}$ \\ 
	  09 ~ {\bf endfor} \\
	  10 ~ $\size{1} \gets \size{2} $ \\
	  11 ~ $\size{2} \gets 1 \\$
	  12 ~ $\FREE_p \gets \mapf{\FREE_p, \size{1}, \size{2}}$ \\
	  13 ~ $\FREE_p \gets \ItStep{\FREE_p, \size{2}}$ 
	}
		\hrule	
		\caption{Algorithm $\ItAlg{\epsilon}$: pseudocode}
		\BB
		\label{fig:IterKK}
\end{figure}

The algorithm  $\ItStepName$   is different from $\algName$ in the following ways. First, all instances of $\ItStepName$
work for $\beta = 3m^2$. Moreover, $\ItStepName$ has a termination
flag in shared memory. This termination flag is initially $0$ and is set to $1$ by any process that decides to terminate.
In the execution of algorithm $\ItStepName$, a process $p$, that in an action
 $\act{compNext}_p$ has $|\FREE_p \setminus \TRY_p| < 3m^2$ , sets the termination flag to $1$,
computes new sets $\FREE_p$ and $\TRY_p$, returns the set $\FREE_p \setminus \TRY_p$ and terminates.
After a process $p$ checks if it is
safe to perform a job, the process also checks the termination flag and if the flag is\,$1$, the process instead of performing
the job, computes new sets $\FREE_p$ and $\TRY_p$, returns the set $\FREE_p \setminus \TRY_p$ and terminates. 
Finally, algorithm $\ItStepName$
takes as inputs the variable $size$ and a set $\SET_1$, such that $|\SET_1| > 3m^2$, and returns the set $\SET_2$ as output. 
$\SET_1$ contains super-jobs of size $size$. In $\ItStepName$, with an action $\act{do}_{p,j}$ process $p$ performs all the jobs of super-job $j$. 
A process $p$ performs as many super-jobs as it can and returns in $\SET_2$ the super-jobs it can verify that no process
will perform. 

In algorithm $\ItAlg{\epsilon}$ we use also the function $\SET_2 = \mapf{\SET_1, {\rm size}_1, {\rm size}_2}$, that takes the set of super-jobs 
$\SET_1$, with super-jobs of size $size_1$ and maps it to a set of super-jobs $\SET_2$ with size $size_2$. 
A job $i$ is always mapped to the same super-job of a specific size and there is no intersection between the jobs in super-jobs of the same size.
%
\subsection{Analysis}
We begin the analysis of algorithm $\ItAlg{\epsilon}$ by showing in 
Theorem\,\ref{thm:italgCorrect} that $\ItAlg{\epsilon}$ solves the at-most-once problem. This is done by first
showing in Lemma\,\ref{lem:italgCorrect1} that algorithm $\ItStepName$ solves the at-most-once problem for the set of all super-jobs of a specific size, 
and then by showing in Lemma\,\ref{lem:italgCorrect2} that there exist no performed super-jobs in any output set $\SET_2$. 
We complete the analysis with Theorem\,\ref{thm:italg}, where we show that 
algorithm $\ItAlg{\epsilon}$ 
has effectiveness $E_{\ItAlg{\epsilon}}(n,m,f) = n-\BigO{m^2 \log n \log m}$ and
 work complexity $W_{\ItAlg{\epsilon}} = \BigO{n + m^{3+\epsilon} \log n}$.

Let the set of all super-jobs of a specific size $d$ be $\SuperSET_d$. All 
invocations of algorithm $\ItStepName$ on sets $\SET_1 \subseteq \SuperSET_d$,
use the matrix $done$ and vector $next$ that correspond to the super-jobs of size 
$d$. Moreover each process $p$ invokes algorithm $\ItStepName$ for a set 
$\SET_1 \subseteq \SuperSET_d$ only once. We have the following lemma.

\begin{lemma}
\label{lem:italgCorrect1}
Algorithm  $\ItStepName$ solves the at-most-once problem for the set $\SuperSET_d$.
\end{lemma}
\begin{proof}
As described above, algorithm  $\ItStepName$ is different from $\algName$ in the
following ways: 
\begin{itemize}
\item Process $p$, on algorithm $\ItStepName$,  has an input set 
$\SET_1 \subseteq \SuperSET_d$ of super-jobs of size $d$ to be performed and outputs a set $\SET_2 \subset \SuperSET_d$ of super-jobs, that have not been performed. Process $p$ initially sets its set $\FREE_p$, equal to $\SET_1$
and proceeds as it would do when executing $\algName$, with the difference that an action $\act{do}_{p,i}$ results in performing all the jobs under super-job $i$. 
Entries in the matrix $done$ and vector $next$ in shared memory correspond to the identifiers of super-jobs of set $\SuperSET_d$. Again after its initialization, entries are only removed from set $\FREE_p$. 

Note that the
main difference caused by this modification, between algorithm $\ItStepName$ and algorithm $\algName$, is that
jobs are replaced by super-jobs, and that the 
initial sets $\FREE_p$ and $\FREE_q$ of processes $p,q$ could be set to different
subsets of set $\SuperSET_d$. This does not affect the correctness of the algorithm, since in any state $\st$ of an execution $\EX$ of algorithm  $\algName$, 
the sets $\FREE_p$ and $\FREE_q$ could be different subsets of the set of all 
jobs $\tset$.

\item Algorithm $\ItStepName$ has a termination
flag in shared memory. The termination flag is initially $0$ and is set to $1$ by any process that decides to terminate. 
As mentioned above, any process that discovers that 
$|\FREE_p \setminus \TRY_p| < 3m^2$ in an action $\act{compNext}_p$, sets the termination flag to $1$, computes new sets $\FREE_p$ and $\TRY_p$, returns the set $\FREE_p \setminus \TRY_p$ and terminates. This modification only affects the sequence of actions during the termination
of a process $p$. Observe process 
$p$ does not perform any super-jobs in that termination sequence. 

Additionally, after a process $p$ checks if it is
safe to perform a super-job, it also checks the termination flag and if the flag 
is\,$1$, the process instead of performing
the super-job, enters the termination sequence,  computing new sets $\FREE_p$ and $\TRY_p$, returning the set $\FREE_p \setminus \TRY_p$ and terminating. 
A process $p$ first checks if it is safe to perform a super-job according to 
algorithm $\algName$ and then checks the flag. Thus this modification only
affects the effectiveness, but not the correctness of the algorithm, since
it could only result in a super-job that was safe to perform not being performed. 

\item Finally all instances of $\ItStepName$
work for $\beta = 3m^2$. This does not affect correctness, since 
Lemma\,\ref{lem:simpleCorrect} holds for any $\beta$.
\end{itemize}

It is easy to
see that none of the modifications described above affect the key arguments in the proof of Lemma\,\ref{lem:simpleCorrect}. Thus with similar arguments as in the 
proof of Lemma\,\ref{lem:simpleCorrect}, we can show 
that there exists no execution of algorithm  $\ItStepName$, where two distinct
actions $\acts = \act{do}_{p,i}$ and $\acts' = \act{do}_{q,i}$ take place for a super-job $i \in \SuperSET_d$ and processes $p,q \in \pset$ ($p$ could be equal to $q$). 
\end{proof}

Next we show that in the output sets of algorithm $\ItStepName$ at a specific
iteration (calls for super-jobs of size $d$), no completed super-jobs are included.
Combined with the previous lemma, this argument will help us establish that
algorithm $\ItAlg{\epsilon}$ solves that at-most-once problem.

\begin{lemma}
\label{lem:italgCorrect2}
There exists no execution $\EX$ of algorithm $\ItStepName$, such that there exists 
action $\act{do}_{q,i} \in \EX$ for some process $q$
and super-job $i$ in the output set $\SET_2 \subset \SuperSET_d$ of some process $p$ 
($p$ could be equal to process $q$).
\end{lemma}
\begin{proof}
As described above, a process $p$ before terminating  algorithm $\ItStepName$,
either sets the flag to $1$ or observes that the flag is set to $1$. The process
$p$ then computes new sets $\FREE_p$ and $\TRY_p$, returns the set $\FREE_p \setminus \TRY_p$ and terminates its execution  of algorithm $\ItStepName$ for input
set $\SET_1 \subseteq \SuperSET_d$ and super-jobs of size $d$. 
Let state $\st$ be the state at which process $p$ terminates, we have that  
$\SET_2 = \st.\FREE_p \setminus \st.\TRY_p$. If $p=q$ and there exists 
action $\acts = \act{do}_{p,i}$  in execution $\EX$  of algorithm $\ItStepName$,
for super-jobs $i \in \SuperSET_d$,
clearly $\acts < \st$, from which we have that $i \notin \st.\FREE_p$ and thus  $i \notin \SET_2$. 

It is easy to see that if $p\neq q$ and $i \in \SET_2$ of process $p$,
there exists no action $\acts = \act{do}_{q,i}$ in execution  $\EX$. 
If $i \in \SET_2$ then $i \in \st.\FREE_p$ and $i \notin \st.\TRY_p$.
Moreover process $p$ either set flag to $1$ or observed that the flag was set,
before computing sets $\st.\FREE_p$ and $\st.\TRY_p$.
If there exists $\acts = \act{do}_{q,i} \in \EX$, for process $q$, it 
must be the case that after process $q$ performed the 
transition $\left(\stt, \act{compNext}_{q},\stt^{'}\right) \ipSym \acts$
(see Definition\,\ref{def:immediatePredecesor} of immediate predecessor),
 it read the flag and found it was equal to $0$. This leads to a contradiction,
since it must be the case that either $i \in \st.\TRY_p$ or 
$i \notin \st.\FREE_p$. 
\end{proof}

We are ready now to show the correctness of algorithm  $\ItAlg{\epsilon}$.

\begin{theorem} 
\label{thm:italgCorrect}
Algorithm  $\ItAlg{\epsilon}$ solves the at-most-once problem.
\end{theorem}
\begin{proof}
From Lemma\,\ref{lem:italgCorrect1} we have that any super-job of a specific 
size $d$ is performed at-most-once (if performed at all) in the execution 
of algorithm $\ItStepName$ for the super-jobs in the set $\SuperSET_d$.
Moreover, from Lemma\,\ref{lem:italgCorrect2} we have that super-jobs in
the output sets of an execution of algorithm $\ItStepName$ for super-jobs
of size $d$, have not been performed. 
Function $\SET_2 = \mapf{\SET_1, {\rm size}_1, {\rm size}_2}$
maps the jobs in the super-jobs of set $\SET_1$, to super-jobs 
in $\SET_2$. A job $i$ is always mapped to the same 
super-job of a specific size $d$ and there is no intersection between the jobs
of the super-jobs in set $\SuperSET_d$.
It is easy to see that there exists no execution of algorithm $\ItAlg{\epsilon}$,
where a job $i$ is performed more than once.
\end{proof}

We complete the analysis of  algorithm $\ItAlg{\epsilon}$ with 
Theorem\,\ref{thm:italg}, which gives upper bounds for the effectiveness and work complexity
of the algorithm.

\begin{theorem} 
\label{thm:italg}
Algorithm $\ItAlg{\epsilon}$ has $W_{\ItAlg{\epsilon}} = \BigO{n + m^{3+\epsilon} \log n}$ 
work complexity and effectiveness $E_{\ItAlg{\epsilon}}(n,m,f) = n-\BigO{m^2 \log n \log m}$.
\end{theorem}
\begin{proof} 
In order to determine the effectiveness and work complexity of algorithm $\ItAlg{\epsilon}$,
we compute the jobs performed by and the work spent in each invocation of 
$\ItStepName$. Moreover
we compute the work that the invocations to the function $\mapf{}$ add. 

The first invocation to function $\mapf{}$ in line $02$ can be completed by process $p$ with work $\BigO{\frac{n}{m \log n \log m} \log n}$,
since process $p$ needs to construct a tree with $\frac{n}{m \log n \log m}$ elements. This contributes for all processes
$\BigO{\frac{n}{\log m}}$ work. 
From Theorem \ref{thm:work} we have that $\ItStepName$ in $03$ has total work 
$\BigO{n + \frac{n}{m \log n \log m} m \log n \log m}=\BigO{n}$, where the first $n$ comes from 
$\act{do}$ actions and the second term from the work complexity of Theorem \ref{thm:work}. Note that
we count $\BigO{1}$ work for each normal job executed by a $\act{do}$ action on a super-job. That means that in 
the invocation of $\ItStepName$ in line $03$, $\act{do}$ actions cost  $m \log n \log m$ work. Moreover from 
Theorem \ref{thm:effectiveness} we have effectiveness $\frac{n}{m \log n \log m} - (3m^2 + m - 2)$
on the super-jobs of size $m \log n \log m$. From the super-jobs not completed, up to $m-1$ may be contained in the $\TRY_p$
sets upon termination in line $03$. Since those super-jobs are not added (and thus are ignored) in the output $\FREE_p$ set
in line $03$, up to $(m-1) m  \log n \log m$ jobs may not be performed by $\ItAlg{\epsilon}$. 
The set $\FREE_p$ returned by algorithm  $\ItStepName$ in line $03$ has no more than
$3m^2 + m - 2$ super-jobs of size $m \log n \log m$.

In each repetition of the loop in lines $04-09$, the $\mapf{}$ function in line $07$ constructs a $\FREE_p$ set with at most 
$\BigO{ m^{2+\epsilon} / \log m}$ elements, which costs $\BigO{m^{2+\epsilon}}$ per process $p$ for a total of
$\BigO{m^{3+\epsilon}}$ work for all processes. Moreover each invocation of  $\ItStepName$ in line $08$ costs 
$\BigO{3m^3 \log n \log m + m^{3+\epsilon} \log m} < \BigO{m^{3+\epsilon}\log n}$ work from Theorem \ref{thm:work}, 
where the term $3m^3 \log n \log m$ is an upper bound
on the work needed for the $\act{do}$ actions on the super-jobs.
From Theorem \ref{thm:effectiveness} we have that each output $\FREE_p$ set in line $08$ has at most $3m^2 + m - 2$ super-jobs.
Moreover from each invocation of $\ItStepName$ in line $08$ at most $m-1$ super-jobs are lost in $\TRY$ sets. Those account
for less than $(m-1) m  \log n \log m$ jobs in each iteration, since the size of the super-jobs in the iterations of the loop in lines $04-09$
is strictly less than   $m  \log n \log m$.

When we leave the loop in lines $04-09$, we have a $\FREE_p$ set with at most $3m^2 + m - 2$ super-jobs of size $\log n \log^{1 + 1/\epsilon} m$,
which means that in line $12$ function $\mapf{}$ will return a set $\FREE_p$ with less than $(3m^2 + m - 2)(\log n \log^{1 + 1/\epsilon} m)$
elements that correspond to jobs and not super-jobs. This costs for all processes a total of 
$\BigO{m^3 \log m \log\log n \log\log m} < \BigO{m^{3+\epsilon} \log n}$ work, since $\epsilon$ is a constant. 
Finally we have that  $\ItStepName$ in line $13$ has from  
Theorem \ref{thm:work} work $\BigO{m^3 \log^2 m \log\log n \log\log m} < \BigO{m^{3+\epsilon} \log n}$ and from 
Theorem \ref{thm:effectiveness} effectiveness $(3m^2 + m - 2)(\log n \log^{1 + 1/\epsilon} m) - (3m^2 + m - 2)$.

If we add up all the work, we have that $W_{\ItAlg{\epsilon}} = \BigO{n + m^{3+\epsilon} \log n}$ since the loop in lines $04-09$
repeats $1+1/\epsilon$ times and $\epsilon$ is a constant. Moreover for the effectiveness, we have that less than or equal to
 $(m-1) m \log n \log m$ jobs will be lost in the $\TRY$ set at line $03$. After that strictly less than $(m-1) m \log n \log m$ jobs will be lost in the
$\TRY$ sets of the iterations of the loop in lines  $04-09$ and fewer than $3m^2 + m - 2$ jobs will be lost from the effectiveness
of the last invocation of $\ItStepName$ in line $13$. Thus we have that $E_{\ItAlg{\epsilon}}(n,m,f) = n-\BigO{m^2 \log n \log m}$.
\end{proof}

\ignore{ 
Algorithm $\ItAlg{\epsilon}$ could be fine tuned so that the jobs lost in the $\TRY$ sets in line $03$ and in the iterations
of loop $04-09$ could be collected before the invocation of $\ItStepName$ in line $13$. This involves modifications in 
$\ItStepName$ so that a process $p$, not only updates about its progress the matrices corresponding to the respective class of super-jobs 
it is preforming at a given execution of $\ItStepName$, but also updates the $done$ matrix corresponding 
to the individual jobs it performs. Moreover an extra announcement space should be reserved in shared memory, 
where for each execution $\ItStepName$ individual jobs to be performed by the processes are announced. 
Finally a process $p$ instead of checking the termination flag only before performing a super-job, it should check  
it before performing each individual job and if the flag is set to $1$, process $p$ should terminate the $\ItStepName$ it is
executing at the given time by collecting the current $\FREE_p \setminus \TRY_p$ . This will allow a process $p$ to collect
the uncompleted jobs from all the $\TRY_q$ sets of previous invocations $\ItStepName$, before executing the $\ItStepName$ in line $13$,
and add them to the respective $\FREE_p$ set after line $12$. All this combined results in effectiveness of
$n - (3m^2 + m - 2)$, without affecting  asymptotically the work complexity.
}
For any $m = \BigO{\sqrt[3+\epsilon]{n/ \log n}}$, algorithm $\ItAlg{\epsilon}$ is work optimal 
and asymptotically effectiveness optimal.
\section{An Asymptotically Optimal Algorithm for the Write-All Problem}
\label{sec:WAItAlg}
\begin{figure}[!ht]
	\hrule
	\FF
	{\scriptsize \noindent{\bf $\WItAlg{\epsilon}$ for process $p$:}} \\
	{\scriptsize
	  00 ~ $\size{1} \gets 1 $ \\
	  01 ~ $\size{2} \gets m\log n \log m \\$
	  02 ~ $\FREE_p \gets \mapf{\tset, \size{1} , \size{2}}$ \\
	  03 ~ $\FREE_p \gets \WItStep{\FREE_p, \size{2}}$\\
	  04 ~ {\bf for}$(i \gets 1, i \leq 1/\epsilon, i++)$\\
	  05 ~ \T $\size{1} \gets \size{2} $ \\
	  06 ~ \T $\size{2} \gets m^{1-i\epsilon} \log n \log^{1+i} m  \\$
	  07 ~ \T $\FREE_p \gets \mapf{\FREE_p, \size{1}, \size{2} }$ \\
	  08 ~ \T $\FREE_p \gets \WItStep{\FREE_p, \size{2}}$ \\ 
	  09 ~ {\bf endfor} \\
	  10 ~ $\size{1} \gets \size{2} $ \\
	  11 ~ $\size{2} \gets 1 \\$
	  12 ~ $\FREE_p \gets \mapf{\FREE_p, \size{1}, \size{2}}$ \\
	  13 ~ $\FREE_p \gets \WItStep{\FREE_p, \size{2}}$ \\
	  14 ~ {\bf for}$(i \in \FREE_p)$\\
	  15 ~ \T $\act{do}_{p,i}$ \\
	  16 ~ {\bf endfor} \\
	}
	  \BB
		\hrule	
		\caption{Algorithm $\WItAlg{\epsilon}$: pseudocode}
		\BB
		\label{fig:WIterKK}
\end{figure}
Based on $\ItAlg{\epsilon}$ we construct algorithm $\WItAlg{\epsilon}$\,Fig.\,\ref{fig:WIterKK}, 
that solves the \emph{Write-All} problem\,\cite{Alex97} with 
work complexity $\mathrm{O}(n + m^{(3+\epsilon)} \log n)$, for any constant $\epsilon > 0$, such that $1 / \epsilon$ is a positive integer.
From Kanellakis and Shvartsman\,\cite{Alex97} the Write-All problem for the shared memory model, consists of: 
``Using $m$ processors write $1$'s to all locations of an array of size $n$.''
The problem assumes that all cells of the array are initialized to $0$.
Algorithm $\WItAlg{\epsilon}$ is different from $\ItAlg{\epsilon}$ in two ways. It uses
a modified version of $\ItStepName$, that instead of returning the $\FREE_p \setminus \TRY_p$ set upon 
termination returns the set $\FREE_p$ instead. Let us name this modified version $\WItStepName$.
Moreover in $\WItAlg{\epsilon}$ after line $13$, process $p$, instead of terminating, executes all jobs in
the set $\FREE_p$. 
Note that since we are interested in the Write-All problem, when process $p$ performs a job $i$ with action 
$\act{do}_{p,i}$, process $p$ just writes $1$, in the $i-$th position of the Write All array $wa[1,\ldots, n]$ in shared memory.

\BB
\begin{theorem}
\label{thm:witalg}
Algorithm $\WItAlg{\epsilon}$ solves the Write-All problem with work complexity $W_{\WItAlg{\epsilon}} = \BigO{n + m^{3+\epsilon} \log n}$.
\end{theorem}
\begin{proof} 
We prove this with similar arguments as in the proof of Theorem \ref{thm:italg}. From Theorem \ref{thm:effectiveness} 
after each invocation of $\WItStepName$ the output set $\FREE_p$ has less than $3m^2 + m -1$ super-jobs.
The difference is that now we do not leave jobs in the $\TRY_p$ sets, since we are not interested in maintaining the at-most-once
property between successive invocations of algorithm $\WItStepName$.
Since after each invocation of $\WItStepName$ the output set $\FREE_p$ has the same upper bound on
super-jobs as in $\ItAlg{\epsilon}$, with similar arguments as in the proof of Theorem \ref{thm:italg}, we have that at line $13$ the total
work performed by all processes is $\BigO{n + m^{3+\epsilon} \log n}$. Moreover  from Theorem \ref{thm:effectiveness} 
the output $\FREE_p$ set in line $p$ has less than $3m^2 + m - 2$ jobs. This gives us for all processes a 
total work of  $\BigO{m^{3}}$ for the loop in lines $14-16$. After the loop in lines 
 $14-16$ all jobs have been performed, since we left no $\TRY$ sets behind, thus algorithm
$\WItAlg{\epsilon}$ solves the Write-All problem with work complexity $W_{\WItAlg{\epsilon}} = \BigO{n + m^{3+\epsilon} \log n}$.
\end{proof}

For any $m = \BigO{\sqrt[3+\epsilon]{n/ \log n}}$, algorithm $\WItAlg{\epsilon}$ is work optimal.
\section{Conclusions}
\label{sec:conclude}
\sk{
We devised and analyzed a deterministic algorithm  for the
at most once problem called $\algName$. 
For $\beta =m$ algorithm $\algName$ has effectiveness  $n-2m+2$, which is asymptotically optimal
for any $m=\mathrm{o}(n)$ and close by an additive factor of $m$ to the effectiveness upper bound $n-m+1$ on all possible algorithms. 
This is a significant improvement over the previous best known deterministic algorithm \cite{KKNS09},
that  achieves asymptotically optimal effectiveness only for $m=\mathrm{O}(1)$. 
With respect to work complexity,  for any constant $\epsilon$ and for 
$m=\BigO{\sqrt[3+\epsilon]{n/ \log n}}$ we demonstrate how to use   $\algName$  with $\beta = 3m^2$, in order to construct
an iterated algorithm $\ItAlg{\epsilon}$, that is work-optimal and asymptotically effectiveness-optimal. Finally
we used algorithm $\ItAlg{\epsilon}$ in order to solve the Write-All problem with work complexity
$\mathrm{O}(n + m^{(3+\epsilon)} \log n)$, for any constant $\epsilon > 0$, which is work optimal 
for $m=\BigO{\sqrt[3+\epsilon]{n/ \log n}}$. Our solution improves on the algorithm of Malewicz~\cite{Malewicz05}
both in terms of the range of processors for which we achieve optimal work and on the fact that
we do not assume test-and-set primitives, but use only atomic read/write shared memory. 
The solution of Kowalski and Shvartsman~\cite{KS08} 
is work optimal for a wider range of processors $m$ than our algorithm, but their algorithm uses a collection
of $q$ permutations with contention $\mathrm{O}(q \log q)$.
Although an efficient polynomial time construction
of permutations with contention $\BigO{q \text{ polylog } q}$ has been 
developed by Kowalski \emph{et al.}\,\cite{Kowalski05},  
 constructing permutations with contention $\BigO{q \log q}$ in polynomial time 
is still an open problem. 
Subsequent to the conference version of this paper\,\cite{KK12}, 
Alistarh \emph{et al.}\,\cite{Dan12} show that there exists a deterministic algorithm 
for the Write-All problem with work $\BigO{n+ m \log^{5} n \log^{2} \max(n,m)}$, by 
derandomizing their randomized solution for the problem. Their solution is so far existential,
while ours explicit.

In terms of open questions there still exists an effectiveness gap between the shown effectiveness
of  $n-2m+2$ of  algorithm $\algName$ and the  known effectiveness bound of $n-m+1$. It would be interesting 
to see if this can be bridged for deterministic algorithms. Moreover,  there is a lack of an upper bound on work complexity,
when the effectiveness of an algorithm approaches the optimal.
Finally it would be interesting to study the existence and efficiency
of algorithms that try to implement at-most-once
semantics in systems with different means of communication,
such as message-passing systems. 
}


\bibliographystyle{abbrv}
\bibliography{Biblio}

\begin{thebibliography}{10}

\bibitem{Afek93}
Y.~Afek, E.~Weisberger, and H.~Weisman.
\newblock A completeness theorem for a class of synchronization objects.
\newblock In {\em Proc. of the 12th annual {ACM} Symp. on Principles of
  Distributed Computing(PODC '93)}, pages 159--170. ACM, 1993.

\bibitem{Dan12}
D.~Alistarh, M.~Bender, S.~Gilbert, and R.~Guerraoui.
\newblock How to allocate tasks asynchronously.
\newblock In {\em Foundations of Computer Science (FOCS), 2012 IEEE 53rd Annual
  Symposium on}, pages 331 --340, Oct. 2012.

\bibitem{AW97}
R.~J. Anderson and H.~Woll.
\newblock Algorithms for the certified write-all problem.
\newblock {\em SIAM J. Computing}, 26(5):1277--1283, 1997.

\bibitem{ABDPR90}
H.~Attiya, A.~Bar-Noy, D.~Dolev, D.~Peleg, and R.~Reischuk.
\newblock Renaming in an asynchronous environment.
\newblock {\em J. ACM}, 37(3):524--548, 1990.

\bibitem{Bayer72}
R.~Bayer.
\newblock Symmetric binary b-trees: Data structure and maintenance algorithms.
\newblock {\em Acta Informatica}, 1:290--306, 1972.

\bibitem{BirNel84}
A.~D. Birrell and B.~J. Nelson.
\newblock Implementing remote procedure calls.
\newblock {\em ACM Trans. Comput. Syst.}, 2(1):39--59, 1984.

\bibitem{Bokal12}
D.~Bokal, B.~Bre\v{s}ar, and J.~Jerebic.
\newblock A generalization of hungarian method and hall's theorem with
  applications in wireless sensor networks.
\newblock {\em Discrete Appl. Math.}, 160(4-5):460--470, Mar. 2012.

\bibitem{CCW95}
S.~Chaudhuri, B.~A. Coan, and J.~L. Welch.
\newblock Using adaptive timeouts to achieve at-most-once message delivery.
\newblock {\em Distrib. Comput.}, 9(3):109--117, 1995.

\bibitem{ChlebusKowal05}
B.~S. Chlebus and D.~R. Kowalski.
\newblock Cooperative asynchronous update of shared memory.
\newblock In {\em STOC}, pages 733--739, 2005.

\bibitem{Czygrinow12}
A.~Czygrinow, M.~Han\'{c}kowiak, E.~Szyma\'{n}ska, and W.~Wawrzyniak.
\newblock Distributed 2-approximation algorithm for the semi-matching problem.
\newblock In {\em Proceedings of the 26th international conference on
  Distributed Computing}, DISC'12, pages 210--222, Berlin, Heidelberg, 2012.
  Springer-Verlag.

\bibitem{GK05}
G.~Di~Crescenzo and A.~Kiayias.
\newblock Asynchronous perfectly secure communication over one-time pads.
\newblock In {\em Proc. of 32nd International Colloquium on Automata, Languages
  and Programming(ICALP '05)}, pages 216--227. Springer, 2005.

\bibitem{Rotem12}
A.~Drucker, F.~Kuhn, and R.~Oshman.
\newblock The communication complexity of distributed task allocation.
\newblock In {\em Proc. of the 31st annual Symp. on Principles of Distributed
  Computing(PODC '12)}, pages 67--76. ACM, 2012.

\bibitem{FLP85}
M.~J. Fischer, N.~A. Lynch, and M.~S. Paterson.
\newblock Impossibility of distributed consensus with one faulty process.
\newblock {\em J. ACM}, 32(2):374--382, 1985.

\bibitem{Fitzi07}
M.~Fitzi, J.~B. Nielsen, and S.~Wolf.
\newblock How to share a key.
\newblock In {\em Allerton Conference on Communication, Control, and Computing
  2007}, 2007.

\bibitem{GS08}
C.~Georgiou and A.~A. Shvartsman.
\newblock {\em Do-All Computing in Distributed Systems: Cooperation in the
  Presence of Adversity}.
\newblock Springer, 2008.

\bibitem{GS11}
C.~Georgiou and A.~A. Shvartsman.
\newblock {\em Cooperative Task-Oriented Computing: Algorithms and Complexity}.
\newblock Synthesis Lectures on Distributed Computing Theory. Morgan {\&}
  Claypool Publishers, 2011.

\bibitem{GL90}
K.~J. Goldman and N.~A. Lynch.
\newblock Modelling shared state in a shared action model.
\newblock In {\em Logic in Computer Science}, pages 450--463, 1990.

\bibitem{JHMV01}
J.~Groote, W.~Hesselink, S.~Mauw, and R.~Vermeulen.
\newblock An algorithm for the asynchronous write-all problem based on process
  collision.
\newblock {\em Distributed Computing}, 14(2):75--81, 2001.

\bibitem{GuibasS78}
L.~J. Guibas and R.~Sedgewick.
\newblock A dichromatic framework for balanced trees.
\newblock In {\em 19th Annual Symposium on Foundations of Computer
  Science(FOCS)}, pages 8--21, 1978.

\bibitem{Harvey06}
N.~J.~A. Harvey, R.~E. Ladner, L.~Lov\'{a}sz, and T.~Tamir.
\newblock Semi-matchings for bipartite graphs and load balancing.
\newblock {\em J. Algorithms}, 59(1):53--78, Apr. 2006.

\bibitem{MH91}
M.~Herlihy.
\newblock Wait-free synchronization.
\newblock {\em ACM Transactions on Programming Languages and Systems},
  13:124--149, 1991.

\bibitem{Hillel2010}
K.~C. Hillel.
\newblock Multi-sided shared coins and randomized set-agreement.
\newblock In {\em Proc. of the 22nd {ACM} Symp. on Parallel Algorithms and
  Architectures ({SPAA}'10)}, pages 60--68, 2010.

\bibitem{Alex97}
P.~C. Kanellakis and A.~A. Shvartsman.
\newblock {\em Fault-Tolerant Parallel Computaion}.
\newblock Kluwer Academic Publishers, 1997.

\bibitem{SCA12}
S.~Kentros, C.~Kari, and A.~Kiayias.
\newblock The strong at-most-once problem.
\newblock In {\em Proc. of 26th International Symp. on Distributed
  Computing(DISC'12)}, pages 390--404, 2012.

\bibitem{KK12}
S.~Kentros and A.~Kiayias.
\newblock Solving the at-most-once problem with nearly optimal effectiveness.
\newblock In {\em ICDCN}, pages 122--137, 2012.

\bibitem{KKNS09}
S.~Kentros, A.~Kiayias, N.~C. Nicolaou, and A.~A. Shvartsman.
\newblock At-most-once semantics in asynchronous shared memory.
\newblock In {\em Proc. of 23rd International Symp. on Distributed
  Computing(DISC'09)}, pages 258--273, 2009.

\bibitem{Kowalski05}
D.~Kowalski, P.~M. Musial, and A.~A. Shvartsman.
\newblock Explicit combinatorial structures for cooperative distributed
  algorithms.
\newblock In {\em Proceedings of the 25th IEEE International Conference on
  Distributed Computing Systems}, ICDCS '05, pages 49--58, Washington, DC, USA,
  2005. IEEE Computer Society.

\bibitem{KS08}
D.~R. Kowalski and A.~A. Shvartsman.
\newblock Writing-all deterministically and optimally using a nontrivial number
  of asynchronous processors.
\newblock {\em ACM Transactions on Algorithms}, 4(3), 2008.

\bibitem{Paxos98}
L.~Lamport.
\newblock The part-time parliament.
\newblock {\em ACM Trans. Comput. Syst.}, 16(2):133--169, 1998.

\bibitem{LLS93}
B.~W. Lampson, N.~A. Lynch, and J.~F. S-Andersen.
\newblock Correctness of at-most-once message delivery protocols.
\newblock In {\em Proc. of the IFIP TC6/WG6.1 6th International Conference on
  Formal Description Techniques(FORTE '93)}, pages 385--400. North-Holland
  Publishing Co., 1994.

\bibitem{LinGan85}
K.-J. Lin and J.~D. Gannon.
\newblock Atomic remote procedure call.
\newblock {\em IEEE Trans. Softw. Eng.}, 11(10):1126--1135, 1985.

\bibitem{Argus88}
B.~Liskov.
\newblock Distributed programming in argus.
\newblock {\em Commun. ACM}, 31(3):300--312, 1988.

\bibitem{LLW91}
B.~Liskov, L.~Shrira, and J.~Wroclawski.
\newblock Efficient at-most-once messages based on synchronized clocks.
\newblock {\em ACM Trans. Comput. Syst.}, 9(2):125--142, 1991.

\bibitem{LT89}
N.~Lynch and M.~Tuttle.
\newblock An introduction to input/output automata.
\newblock {\em CWI-Quarterly}, pages 219--246, 1989.

\bibitem{Lynch1996}
N.~A. Lynch.
\newblock {\em Distributed Algorithms}.
\newblock Morgan Kaufmann Publishers, 1996.

\bibitem{Malewicz05}
G.~Malewicz.
\newblock A work-optimal deterministic algorithm for the certified write-all
  problem with a nontrivial number of asynchronous processors.
\newblock {\em SIAM J. Comput.}, 34(4):993--1024, 2005.

\bibitem{Spector82}
A.~Z. Spector.
\newblock Performing remote operations efficiently on a local computer network.
\newblock {\em Commun. ACM}, 25(4):246--260, 1982.

\bibitem{Watson89}
R.~W. Watson.
\newblock The delta-t transport protocol: Features and experience.
\newblock In {\em Proc. of the 14th Conf. on Local Computer Networks}, pages
  399--407, 1989.

\end{thebibliography}

\end{document}